\documentclass[prb,aps,twocolumn,superscriptaddress]{revtex4-2}
\usepackage{bm}
\usepackage[colorlinks=true,linkcolor=blue,citecolor=blue]{hyperref}
\usepackage{times}
\usepackage{amsmath}
\usepackage{amssymb}
\usepackage{amsthm}
\usepackage{amsfonts}
\usepackage{enumerate}
\usepackage{latexsym}
\usepackage{ifpdf}
\newcommand{\beq}{\begin{equation}}
\newcommand{\eeq}{\end{equation}}
\usepackage{graphicx}
\usepackage{makeidx}
\usepackage{color}

\usepackage[version=4]{mhchem}
\usepackage{textcomp,mathcomp}
\usepackage{verbatim}

\begin{document}

\title{When electrons meet ferroelastic domain walls in Strontium Titanate}

 \author {Shashank Kumar Ojha}
 \email{so37@rice.edu}
\altaffiliation{Contributed equally}
\affiliation{Department of Physics, Indian Institute of Science, Bengaluru 560012, India}
\affiliation  {Rice Advanced Materials Institute, Rice University, Houston, Texas 77005, USA}
\affiliation  {Materials Science and NanoEngineering, Rice University, Houston, Texas 77005, USA}
\affiliation  {Department of Physics, Rice University, Houston, Texas 77005, USA}
\author {Jyotirmay Maity}
\altaffiliation{Contributed equally}
\affiliation {Department of Physics, Indian Institute of Science, Bengaluru 560012, India}
\author {Srimanta Middey}
\email{smiddey@iisc.ac.in}
\affiliation{Department of Physics, Indian Institute of Science, Bengaluru 560012, India}

\begin{abstract}
Strontium titanate (SrTiO$_3$), famously described by Nobel laureate K. A.  M\"uller as the “drosophila of solid-state physics", has been extensively investigated over the last seventy five years for its intricate coupling of structural, electronic, and dielectric properties and continues to serve as a foundational platform for advancing oxide electronics. In its pristine form, SrTiO$_3$  exhibits quantum paraelectric behavior below 35 K and undergoes an antiferrodistortive phase transition near 105 K. This transition generates ferroelastic twin domains separated by a dense network of domain walls, which function as nanoscale structural defects with far-reaching consequences.
While the static influence of ferroelastic domain walls on carrier transport in electron-doped SrTiO$_3$ is well established, recent experimental results show that the emergence of polarity at these walls, combined with strain fields and inherent quantum fluctuations, induces correlated dynamical phenomena such as glass-like relaxations of electrons and memory effects. In this review, we highlight these recent advances, focusing on the subtle interplay between the emergence of nanoscale polar order, quantum fluctuations, and long-range strain fields. We propose that understanding charge carrier dynamics in the background of these complex ferroelastic domain wall landscapes offers a new paradigm for exploring electronic transport in the presence of local polar order and quantum fluctuations, with broad implications for correlated oxides.

\end{abstract}

\maketitle

\section{Introduction}

Condensed matter systems that host multiple coexisting or competing orders have attracted significant attention in recent years for discovering complex and often unexpected phenomena~\cite{Fradkin:2015p457,middey:2016p305,Fernandes:2019p133,Junquera:2023p025001,mostovoy:2024p18,yu:2025pe07070,Nataf:2020p634,Wang:2019p61,Armitage:2010p2421,Keimer:2015p179,Tokura:2017p2017}. The intricate interplay among these distinct yet coupled orders gives rise to rich phase diagrams, unconventional excitations and transitions, enhanced sensitivity to external perturbations, and the emergence of collective behaviors, features absent in systems dominated by a single order parameter.
Understanding such interactions not only addresses fundamental questions in condensed matter physics but also opens avenues for designing materials with enhanced or entirely new functionalities. Continuous advances in experimental techniques have further expanded this frontier, enabling the discovery of new phenomena even in long-studied compounds, thereby motivating a re-examination with fresh perspectives.
In this context, we spotlight SrTiO$_3$
 (STO), as a prototypical system that has captivated the condensed matter physics and materials science community for more than seven decades, offering crucial insights and continuing to provide exciting opportunities for exploring complex physical phenomena~\cite{Bussmann:2024p3}.

\begin{figure*} [t!]
	\vspace{-1pt}
	\hspace{0pt}
    \centering
	\includegraphics[width=1\linewidth] {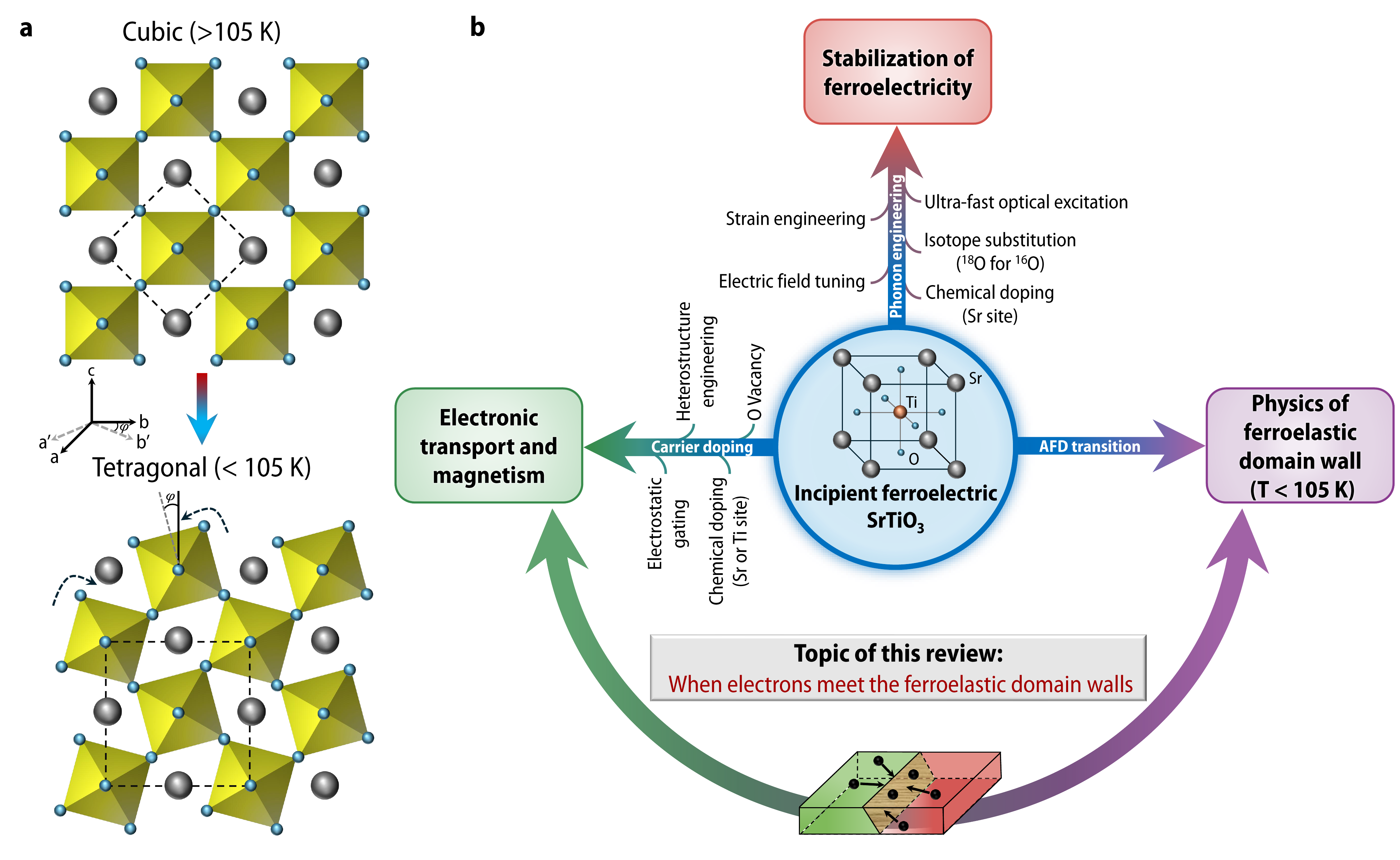}
	\caption{\textbf{a}. Schematic illustrating the temperature-driven  AFD structural transition of STO from the high-temperature cubic phase (above 105 K) to the low-temperature tetragonal phase (below 105 K). This transition is driven by the rotation of TiO$_6$ octahedra, resulting in the formation of a complex network of ferroelastic twin domains (discussed later in detail). \textbf{b}. Overview of the major research frontiers in incipient ferroelectric STO. Primarily, research has focused on two major fields: stabilizing ferroelectricity via external perturbation (e.g., chemical doping, isotope substitution, strain engineering, and ultrafast optical excitations, etc.) and exploring emergent electronic transport and magnetism through carrier doping. More recently, the intrinsic physics of ferroelastic domain walls has emerged as a distinct field of interest. The central theme of this review (bottom curved arrow) addresses the critical intersection of these areas: understanding how the ferroelastic domain wall actively affects and controls the electronic transport properties in doped STO. \label{Fig0}}
\end{figure*}

The synthesis of bulk STO was first reported in 1950 by J. K. Hulm~\cite{Hulm:1950p1184}. It has a perfect cubic perovskite structure at room temperature (Fig.~\ref{Fig0}a) and is a wide band gap insulator with an indirect band gap of 3.2 eV. Its high refractive index ($\sim$2.4) and low reciprocal relative dispersion ($\sim$13) initially highlighted STO as a promising optical material, while defect-induced color tuning even suggested potential use as a gemstone. Recognizing these attributes, Leon Marker patented the synthesis of STO crystal in 1953~\cite{STOpatent}. In 1958, K. A. M\"uller first reported a cubic-to-tetragonal structural phase transition at approximately $T_{AFD}$$\approx$ 105 K upon cooling~\cite{Muller:1958thesis}.
This transition was subsequently identified as an antiferrodistortive (AFD) instability, characterized by staggered (antiphase) rotations of TiO$_6$ octahedra around the $c$-axis [Fig.~\ref{Fig0}a] ~\cite{Cowley:1964pA981,Muller:1968p814,fleury:1968p16,shirane:1969p858,Cowley:1969p181}. Further, this ferroelastic transition results in a dense network of ferroelastic twin domains with distinct walls separating each group of
domains (discussed later). Advances in understanding this transition up to 1996 were comprehensively reviewed by Cowley \textit{et al.}~\cite{cowley:1996p2799}, and we will revisit some of its fundamental aspects later in this article.  More recently, Collignon \textit{et al.}~\cite{Collignon:2019p031218} summarized how chemical substitution and oxygen vacancies can be used to manipulate the AFD transition.

 Another key attribute of STO lies in its dielectric response. At room temperature, STO exhibits a remarkably large dielectric constant ($\epsilon$ $\sim$ 250–300), far exceeding that of widely used compounds such as SiO$_2$ (3.9), Si$_3$N$_4$
 (7), Al$_2$O$_3$ (9), TiO$_2$ (80), and HfO$_2$ (25) etc~\cite{Locquet:2006p051610}. Upon cooling, the dielectric constant increases substantially ($\epsilon$ $\sim$ 10,000)~\cite{Neville:1972p2124,Viana:1994p601,Sakudo:1971p851}; however, STO does not undergo a conventional ferroelectric transition due to the suppression of long-range order by quantum fluctuations. Instead, below 35 K, STO enters a quantum paraelectric phase~\cite{Muller:1979p3593}, which we revisit later in this review. Considerable efforts have been devoted to stabilizing a ferroelectric phase through various approaches, as summarized in Fig.~\ref{Fig0}b~\cite{Fleury:1968p613,Hemberger:1995p13159,Bednorz:1984p2289,Itoh:1999p3540,Haeni:2004p758,Lee:2015p1314,Nova:2019p1075}.

The exceptionally large dielectric constant also enables a metallic phase in STO at extremely dilute carrier concentrations, consistent with Mott’s criterion for the metal–semiconductor transition~\cite{Frederikse:1963pA442,mott:2004metal}. This phenomenon was recognized very early, with oxygen-deficient STO identified as the first oxide superconductor~\cite{Schooley:1964p474}. Parallel strategies involving chemical doping at either the Sr or Ti sites have been explored to tune metallic transport. Despite these pioneering observations, the microscopic mechanism underlying unconventional superconductivity in STO remains unresolved, and we refer readers to existing comprehensive reviews for detailed discussions~\cite{Collignon:2019p031218,GASTIASORO:2020p168107}. Over the past twenty years, advances in thin-film growth techniques have further established STO-based oxide two-dimensional electron gases (2DEGs) as a major research theme, as it hosts a variety of collective behavior including tunable Rashba spin-orbit coupling, orbital reconstruction, metal-insulator transition, superconductivity, ferromagnetism, etc.~\cite{Ohtomo:2004p423,Thiel:2006p1942,Reyren:2007p1196,Caviglia:2008p624,Caviglia:2010p126803,Bert:2011p767,Salluzzo:2013p2333}. Numerous review articles provide thorough accounts of this progress~\cite{Pai:2018p036503,Huijben:2009p1665,Gariglio:2016p060701,Stemmer:2014p151,trier:2018p293002,Chen:2010p2881,Steegemans:2025p2119,Chandra:2017p112502}.

In parallel, several studies over the past decade have uncovered the presence of local polar order confined to ferroelastic domain walls in insulating STO, even though the bulk domains themselves remain non-polar~\cite{Zubko:2007p1667601,Scott:2012p187601,Salje:2013p247603,Kalisky:2013p1091,Honig:2013p1112,Frenkel:2017p1203}.
 Well below the AFD transition, the coupling between these localized polar regions, along with the inherent global ferroelectric fluctuations and strain fields from the underlying ferroelastic order, gives rise to emergent collective domain wall phases, such as the quantum domain glass and quantum domain solid~\cite{Pesquera:2018p235701,Kustov:2020p016801,Fauque:2022pL140301,Casals:2019p032025}. Such phases highlight the interplay between strain and polarization, which can generate glassy and nonequilibrium phenomena in a system that is otherwise considered structurally simple and nonpolar.

The collective behavior of domain walls becomes even more intriguing and is the central focus of this review when mobile charge carriers are also present.
Conventionally, one might expect that the free carriers in a metal would effectively screen the local electric dipoles, thereby quenching this local polar order~\cite{Bhowal:2023p53}.

However, a large amount of recent experimental works have demonstrated the opposite: the polar order at the domain walls not only persists in the doped regime but also profoundly influences electronic transport. In this review, we discuss these developments to provide a comprehensive picture of how ferroelastic domain walls influence charge transport in electron-doped STO. We will connect the static influence of these domain walls to their more recently uncovered dynamical aspects, arguing that these domain walls are not passive scatterers but active, dynamic entities that are essential for understanding the material's complex electronic properties. The article is structured as follows: we begin by reviewing the relevant order parameters in the bulk of STO  and their role in understanding the emergence of additional polar order at domain walls. We then describe the static influence of domain walls on electronic transport in doped STO, followed by a discussion on collective phases of domain walls and their relevance in understanding the dynamical aspects of electronic relaxation. We conclude by placing these findings in a broader context, drawing connections to related fields, including metal-insulator
transitions, electron glass, quantum critical metals, and the physics of polar metals and superconductors. We also emphasize that this review highlights only the recent advances in the domain-wall physics of STO, with an emphasis on emerging evidence for collective behavior and glassy relaxation dynamics. Given the extensive body of prior work on STO, we do not aim for completeness; instead, we focus on selected results that elucidate the key physical mechanisms and experimental approaches underlying these phenomena. We apologize for any omissions.

\begin{figure*} [t!]
	\vspace{-1pt}
	\hspace{0pt}
\centering
\includegraphics[width=0.8\linewidth] {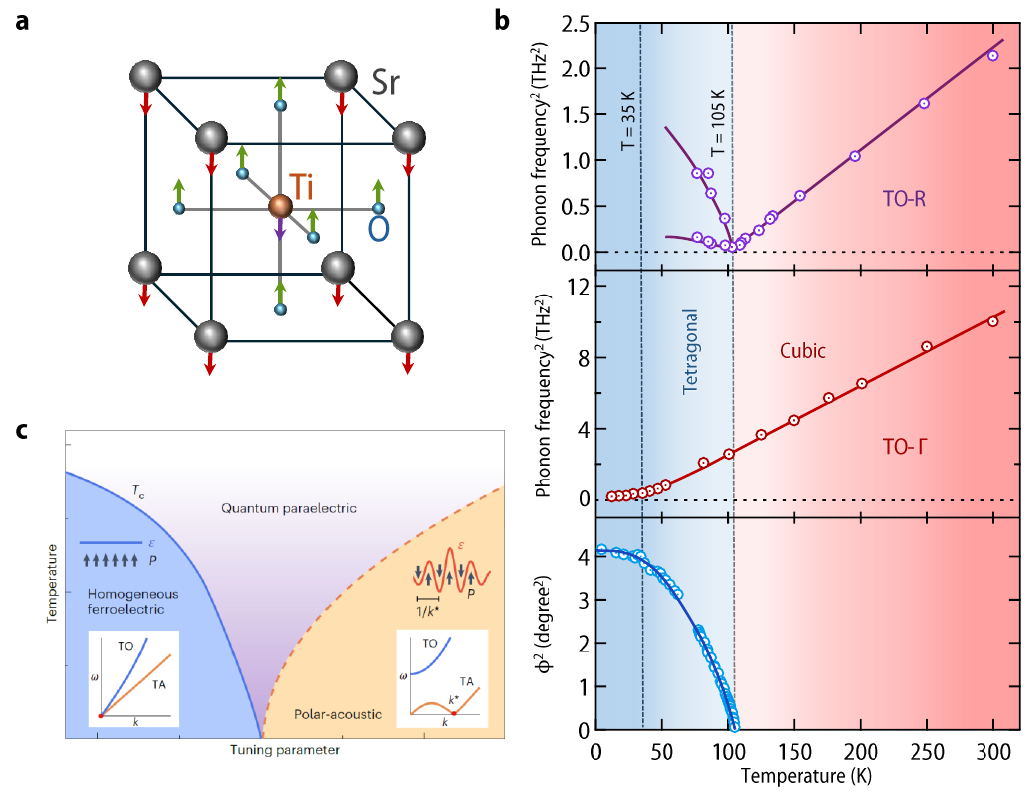}
	\caption{\textbf{a}. Unit cell schematic of STO illustrating transverse optical (TO) soft-mode. \textbf{b}. Temperature dependence of the soft-mode frequencies at the R point, $\Gamma$ point, and antiphase rotation angle of TiO$_6$ octahedra ($\phi$). The R-point mode (TO-R) softens at 105 K, marking the cubic to tetragonal structural transition. Below this transition, the lowering of crystal symmetry lifts the triple degeneracy of the R-point soft mode, causing it to split into two distinct branches: a singlet corresponding to rotation about the unique $c$-axis and a doublet corresponding to rotations about the equivalent $a$- and $b$-axes~\cite{Cowley:1969p181}. Both branches harden with decreasing temperature. In comparison, the $\Gamma$-point TO mode (TO-$\Gamma$) continues to soften upon cooling but tends to saturate below $\sim$ 35 K due to quantum fluctuations, signifying the emergence of the quantum paraelectric state in STO. The bottom panel shows the temperature dependence of the antiphase rotation angle of TiO$_6$ octahedra ($\phi$) in the tetragonal phase. This panel has been reproduced from references.~\cite{Cowley:1969p181,Sirenko:2000p373,hayward:1999p501}. \textbf{c}. The schematic phase diagram summarizes the nearby phases around the quantum paraelectric regime. The blue and orange regions denote the ferroelectric phase and a polar-acoustic regime, respectively, which can be accessed via tuning parameters such as flexoelectric coupling. The insets on the left and right depict representative phonon dispersions near the transitions to the FE and polar-acoustic phases, respectively. Panel \textbf{c} is taken from ref.~\cite{Orenstein:2025p961}. \label{Fig1} }
\end{figure*}

\section{Relevant Order parameters and coupling between phonons and/or order parameters in the bulk of SrTiO$_3$ }

Since much of our understanding of domain wall physics in STO stems from the behavior of bulk order parameters, phonons, and their couplings, this section briefly discusses the foundational concepts that will serve as the basic ingredients for discussing emergent polar order at the twin walls and their collective dynamics.

\subsection{Structural order}

In the mid-20$\textsuperscript{th}$ century, the condensed matter physics community confronted a foundational question: why do certain crystals undergo symmetry changes as temperature varies~\cite{bruce:1981structural,fujimoto:2005physics,toledano:1987landau,cowley:1980p1}. For magnetic systems, Landau’s phenomenological theory of phase transitions~\cite{toledano:1987landau} provided a rigorous framework by expressing the free energy as a function of an order parameter, thereby predicting whether transitions are continuous or first-order. However, in structural phase transitions—where atomic positions rearrange to new equilibrium configurations—the microscopic origin of the instability remained elusive. Although Landau theory captured the symmetry changes and critical scaling, it did not clarify which lattice degrees of freedom were responsible for the emergent instability, i.e., what in the lattice was “going soft”.
A major conceptual advancement emerged in the late 1950s when Cochran and Anderson introduced the soft mode theory~\cite{Cochran:1959p412,cochran:1960p387,anderson:1997concepts,Scott:1974p83,venkataraman:1979p129,Shirane:1970p155} of structural phase transition. This framework posits that a structural phase transition is driven by the progressive softening of a specific normal mode of lattice vibration—a phonon—whose frequency decreases with temperature and becomes zero at the critical temperature. At this point, the restoring force opposing the corresponding lattice distortion vanishes, enabling the crystal to spontaneously lower its symmetry and adopt a new ground state.

STO emerged as the canonical system to validate this theory. While it's AFD transition was long known, Cochran’s proposal galvanized focused experimental efforts to identify the associated soft phonon mode (Fig. \ref{Fig1}a-b). In the late 1960s and early 1970s, detailed Raman spectroscopy and inelastic neutron scattering measurements demonstrated~\cite{Cowley:1964pA981,fleury:1968p16,shirane:1969p858,cowley:1996p2799,hayward:1999p501} that the frequency ($\omega$) of the TO phonon at the Brillouin zone $R$-point (TO-R) decreased systematically as $T$ $\rightarrow$ $T_{AFD}$ from above, ultimately vanishing at the transition (Fig. \ref{Fig1}b). Below $T_{AFD}$, the phonon frequency hardens again~\cite{Cowley:1969p181,fleury:1968p16}, reflecting stabilization of the distorted tetragonal phase. The distortion corresponds to staggered rotations of the TiO$_6$ octahedra as discussed earlier, with the rotation angle ($\phi$) serving as the structural order parameter (Fig.\ref{Fig0}a, \ref{Fig1}b). This landmark observation provided the first direct microscopic confirmation of the soft-mode paradigm as the fundamental driver of a displacive structural phase transition. The convergence of Landau phenomenology and lattice dynamics was elegantly captured by the relation:

\begin{equation}
    \omega^{2}(T) \sim (T - T_{\mathrm{AFD}}),
\end{equation}

where the square of the soft phonon frequency corresponds to the curvature of the Landau free energy near the instability~\cite{Scott:1974p83,venkataraman:1979p129}. This unification enabled predictive insights into structural instabilities through identification and tracking of critical phonon modes. The STO case has since become a paradigm for a broad class of structural transitions: the zone-center soft mode driving ferroelectricity in BaTiO$_3$, the zone-boundary soft mode in the $\alpha$-$\beta$ quartz transition, and finite-wavevector soft modes underlying charge density wave order~\cite{Scott:1974p83,wilson:1975p117,gruner:2018density}.

\subsection{Ferroelastic Order}

A direct macroscopic consequence of this AFD transition is the emergence of ferroelastic order. The spontaneous selection of a unique rotation axis (the tetragonal $c$-axis)
lowers the crystal symmetry and breaks the cubic degeneracy, producing a
ferroelastic distortion. The ferroelastic order parameter ($Q_{\mathrm{F}}$)  is conventionally defined as:

\begin{equation}
    Q_{\mathrm{F}} = \varepsilon_{xx} - \varepsilon_{yy},
\end{equation}

where $\varepsilon$$_{ij}$ are components of the strain tensor. In the
high-temperature cubic phase, $Q_{\mathrm{F}}$ = 0 because the in-plane lattice
constants are equal ($a$ = $b$ = $c$). In the tetragonal phase, the octahedral
rotation axis causes $c$ $\neq$ $a$ (and $a$ = $b$), introducing anisotropic strain
so that $Q_{\mathrm{F}}$ $\neq$ 0. Landau–Ginzburg–Devonshire free-energy treatments capture this effect through a
rotostrictive coupling term~\cite{Zubko:2007p1667601,morozovska:2012p6,Morozovska:2012p094107,Salje:2011p275901,hayward:1999p501,Casals:2018p217601,Buckley:1999p1653,Schranz:2022p161775} of the form:

\begin{equation}
F_{\mathrm{coupling}} \sim \phi^2\, Q_{\mathrm{F}},
\end{equation}

where $\phi$ is the AFD rotation amplitude as discussed earlier. This term ensures that the onset of
octahedral rotation is intrinsically accompanied by ferroelastic strain thereby linking ferroelastic order with the onset of structural phase transition.

\subsection{Suppressed Ferroelectric Order}

STO exhibits a second lattice instability at the Brillouin zone center ($\Gamma$ point) associated with the TO $\Gamma_{15}$ phonon mode (TO-$\Gamma$). In a purely classical picture, the progressive softening of this polar phonon upon cooling would drive a displacive ferroelectric transition at low temperature, with the Ti$^{4+}$ ion shifting off-center within the oxygen octahedron to break inversion symmetry (Fig. \ref{Fig1}a). However, stoichiometric STO remains paraelectric down to the lowest measurable temperatures. This suppression of long-range ferroelectric order arises from strong quantum fluctuations of the lattice, which preempt the condensation of the soft mode~\cite{Muller:1979p3593,Fujishita:2016p074703,Rowley:2014p367}. In the quantum paraelectric regime, the TO-$\Gamma$ mode remains underdamped but saturates at a finite frequency when the amplitude of the TO-$\Gamma$ mode reaches close to the amplitude of the quantum fluctuation, thereby never reaching the instability threshold (Fig. \ref{Fig1}b). This reflects in the dielectric constant which grows to exceptionally large values on cooling, following a Barrett-type temperature dependence rather than the classical Curie--Weiss law, and saturates below $\sim$ 4 K~\cite{Muller:1979p3593}. STO thus stands as the canonical example of a quantum paraelectric, poised on the verge of a ferroelectric quantum critical point (QCP)~\cite{Rowley:2014p367}. The proximity to the QCP renders STO highly sensitive to external perturbations such as isotopic substitution (e.g., $^{18}$O), uniaxial stress, or chemical doping, each of which can induce a ferroelectric ground state~\cite{Itoh:1999p3540,Lee:2015p1314,Haeni:2004p758,Nova:2019p1075,Bednorz:1984p2289}.

\vspace{0.25cm}

\subsection{Coupling between phonons and/or order parameters in the bulk}

\begin{figure*} [t!]
	\vspace{-1pt}
	\hspace{0pt}
	\includegraphics[width=\linewidth] {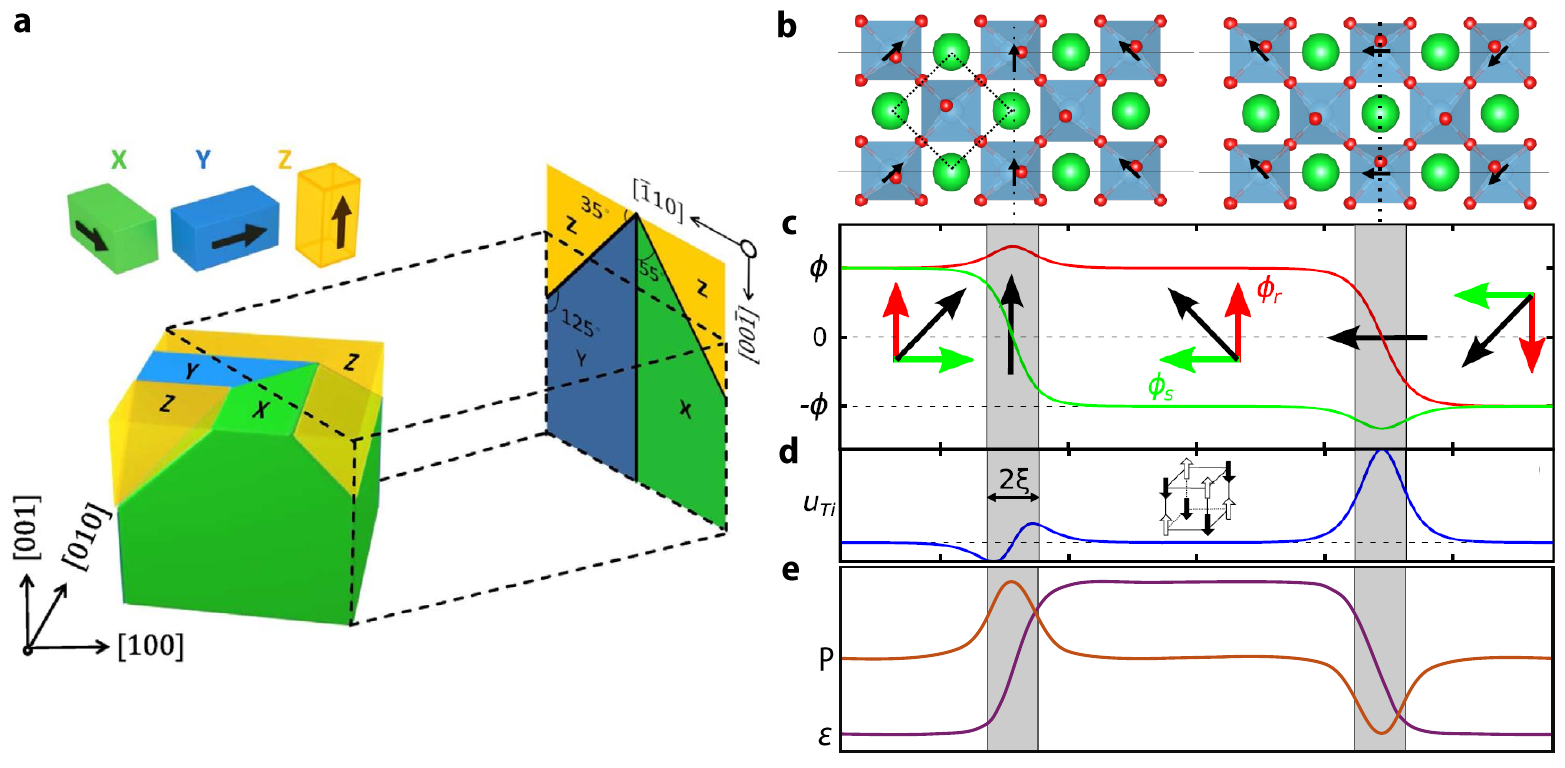}
	\caption{\textbf{a}. A schematic showing structural twin domains of STO below 105 K. Three types of domains with $c$ axis along [100], [010], and [001] have been marked with $X$ (green), $Y$ (blue), and $Z$ (yellow), respectively. The right panel shows the domain geometry in the $(110)$ cross-sectional plane.  The twin boundaries separating $Z$ domains from the $X$ (or $Y$) domains form the characteristic angles of either $55^{\circ}$ ($125^{\circ}$) or $145^{\circ}$ ($35^{\circ}$) in the $(110)$ plane. The wall separating the $X$ and $Y$ domains is aligned along the $[001]$ axis in the $(110)$ plane. This panel is taken from the ref.~\cite{Harsan:2016p257601} \textbf{b}. Atomic representation of two types of domain walls distinguished by the sign reversal of the antiferrodistortive (AFD) pseudovector components $\phi$$_r$ or $\phi$$_s$ corresponding to “head-to-head” (HH, left) and “head-to-tail” (HT, right) configurations. The black arrows indicate the
local tilt vector. Panels \textbf{c} - \textbf{e}. show how different order parameters change spatially across the domain wall. \textbf{c}. Spatial variation of the pseudovector components $\phi$$_r$ and $\phi$$_s$ across two domain walls, with the shaded region denoting the nominal wall width (2$\xi$). \textbf{d}. Amplitude of the antiferroelectric Ti displacement mode, $u_{Ti}$ across the domain walls. The inset illustrates the antiferroelectric character of the Ti displacements. \textbf{e}. Spatial variation of the spontaneous strain ($\epsilon$) and the in-plane polarization ($P$) across the domain walls. Panels \textbf{b} - \textbf{d} are taken from ref.~\cite{Schiaffino:2017p137601}\label{Fig2} }
\end{figure*}

\begin{figure*} [t!]
	\vspace{-1pt}
	\hspace{0pt}
	\includegraphics[width=\linewidth] {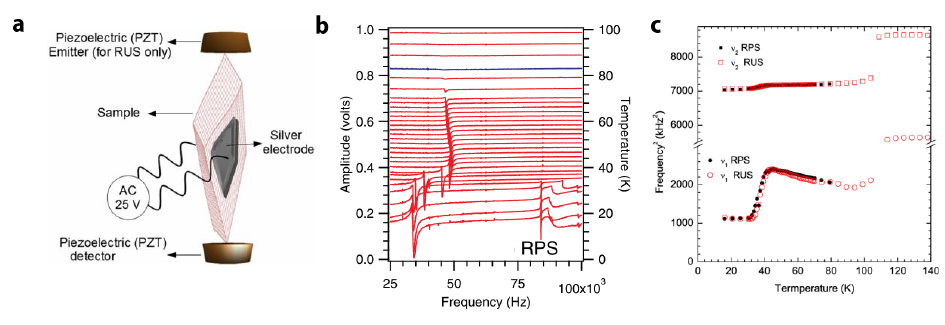}
	\caption{\textbf{a}. Schematic of the experimental configuration for resonant piezoelectric spectroscopy (RPS). The same setup can be adapted for resonant ultrasound spectroscopy (RUS) by applying the ac voltage to the upper piezoelectric transducer instead of across the sample electrodes. The applied ac voltage in both cases was 25 V. \textbf{b}. Low-temperature RPS spectra of STO recorded between 25 kHz and 100 kHz. The spectrum shown in blue was collected  at $\sim$ 80 K. Disappearance of two mechanical resonances in RPS spectra near 40 kHz ($\nu$$_1$) and 85 kHz ($\nu$$_2$) with increasing temperature indicates softening of elastic modes. \textbf{c}. Temperature evolution of the resonance frequencies $\nu$$_1$ and $\nu$$_2$ obtained from RPS and RUS. The close agreement between the two methods confirms that the Fano-like features observed in RPS originate from intrinsic mechanical resonances of STO. Crucially, since bulk STO is centrosymmetric and non-piezoelectric, it cannot normally be driven into mechanical resonance by an electric field. Therefore, the appearance of resonances implies that the domain walls themselves are polar; the applied voltage causes the walls to oscillate, generating the strain fields that drive the sample’s mechanical modes. The polarity at the domain wall develops at 80 K and further enhances below 40 K ~\cite{Scott:2012p187601,Salje:2013p247603}. This figure is taken from ref.~\cite{Salje:2013p247603}\label{Fig3} }
\end{figure*}

Thus far, we have discussed the presence of various distinct order parameters in bulk STO. However, these order parameters are not independent; they are intrinsically coupled to one another. For example, the structural and suppressed ferroelectric order parameters are strongly intertwined~\cite{Sai:2000p13942,Lasota:1997p109}. A prime example is the coupling between the AFD octahedral rotation mode (TO-R) and the TO-$\Gamma$ mode. First-principles studies~\cite{Sai:2000p13942} revealed that these two lattice instabilities are mutually antagonistic: condensation of the TO-R mode in the tetragonal phase renormalizes the TO-$\Gamma$ mode frequencies anisotropically, thereby suppressing the dielectric response. In turn, path-integral simulations~\cite{Vacher:1992p45,Zhong:1996p5047} demonstrate that quantum fluctuation not only suppresses long-range ferroelectric order but also shifts the AFD transition temperature, thereby altering the delicate balance between these competing instabilities. Beyond this direct interaction, both order parameters couple strongly to the ferroelastic order via strain fields. For example, rotostriction links the octahedral tilts (TO-R mode) to the elastic degrees of freedom leading to elastic anomalies in the vicinity of the transition~\cite{Bussmann:2002p141}, while electrostriction couples the TO-$\Gamma$ polar mode to the strain~\cite{morozovska:2012p6}.

Interestingly, dielectric and neutron-scattering experiments indicate the presence of significant finite-wavevector polar-acoustic fluctuations in the quantum paraelectric regime~\cite{muller:1991p277,Vacher:1992p45,coak:2020p12707,Fauque:2022pL140301}. These fluctuations may signal an additional instability distinct from the uniform ferroelectric state-possibly toward a phase with spatially modulated polarization. Such a state could arise from polarization density waves (Fig. \ref{Fig1}c) driven by the coupling between transverse acoustic (TA) phonons and the polar order parameter (from suppressed ferroelectric order) via electrostriction, leading to nanoscale lattice and polarization fluctuations~\cite{Orenstein:2025p961}.

Having discussed the couplings between different phonons and their associated order parameters, we now turn to how these coupled orders can give rise to emergent polarity at twin walls, with intriguing consequences, particularly under electron doping. We also note that, while the above manifestations of order-parameter coupling have primarily been studied in pristine STO, these interactions must be carefully considered when interpreting the emergent behaviors observed in electronic transport upon doping, some of which will be addressed in the upcoming sections.

\section{Emergent polar order at the twin wall and its coupling with other prevailing orders in the bulk}

Domain walls in ferroic materials are increasingly recognized as hosts of functionalities that are absent in the bulk lattice~\cite{Salje:1990p111,Salje:2012p265,Catalan:2012p119}. Two microscopic mechanisms underlie these effects. First, the primary order parameter often vanishes or reconfigures at the wall, thereby activating latent secondary instabilities that are suppressed in the homogeneous phase. Second, the rapid spatial variation of structural distortions generates large gradients, enhancing gradient couplings such as flexoelectricity. Together, these effects frequently give rise to emergent functionalities at the nanoscale boundaries between domains, forming the foundation of the rapidly growing field of domain wall-based nanoelectronics~\cite{Catalan:2012p119,McCluskey:2025p050901,Sharma:2019p2927,Wang:2022p3255,Vermeulen:2024p696,Ghosal:2025parXiv}.

\begin{figure*} [t!]
\centering
	\vspace{-1pt}
	\hspace{0pt}
	\includegraphics[width=0.6\textwidth] {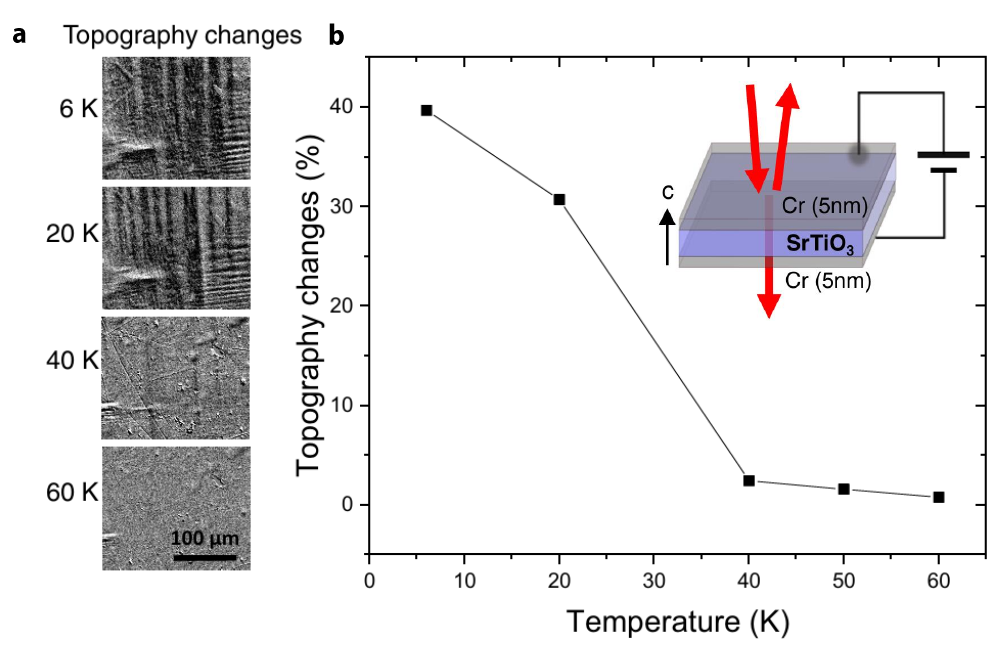}
	\caption{\textbf{a}. Electric-field-induced topographical changes obtained by subtracting reflection images taken at 400 V/mm from 0 V/mm. Pronounced contrast at 6 K and 20 K indicates active twin motion, whereas changes diminish at 40 K and vanish by 60 K. \textbf{b}. Temperature dependence of the topography changes, quantified as the normalized mean intensity of the subtracted images shown in panel \textbf{a}. The magnitude of topography changes decreases sharply above 40 K, indicating a rapid reduction in twin-wall mobility with increasing temperature. The inset shows the schematic of the experimental geometry used for electric field-dependent optical imaging. This figure is taken from ref.~\cite{Casals:2019p032025}\label{Fig4} }
\end{figure*}

In the context of STO, ferroelastic twin walls (Fig.~\ref{Fig2}a) have attracted significant attention in recent years. Above the AFD transition, STO is cubic, allowing three degenerate choices for the tilt axis upon entering the tetragonal phase. In the tetragonal phase, the octahedral tilt pattern is described by $a^0$$a^0$$c^-$ in Glazer notation~\cite{Glazer:1972p3384}, with neighboring domains differing by a 90$^\circ$ reorientation of the tilt axis. The interface between such domains, the twin wall, accommodates this rotation. While the bulk tetragonal domains are nonpolar, the twin boundary locally breaks inversion symmetry, providing a natural site for emergent polarity~\cite{van:2012p523}. One of the earliest experimental demonstrations of this emergent polarity at twin walls was reported by Salje $\textit{et al.}$ through the observation of electromechanical resonances~\cite{Salje:2013p247603} confined to twin domain walls at low temperatures (see Fig.~\ref{Fig3} for a detailed explanation). They found that polarity at the domain wall starts to develop below 80 K and further enhances below 40 K~\cite{Scott:2012p187601,Salje:2013p247603}.

Theoretical frameworks have long emphasized the flexoelectric mechanism as the primary source of polarity at ferroelastic twin walls. Because these walls separate domains of distinct strain states, the resulting strain gradient inherently induces a local polarization (since flexoelectricity is a universal property of all the insulating crystals)~\cite{Morozovska:2012p094107,morozovska:2012p6}. Early phenomenological models invoking this mechanism have since been substantiated by atomistic simulations~\cite{Salje:2016p024114}. However, more recent theoretical analyses~\cite{Schiaffino:2017p137601} have revealed that the microscopic origin of polarity at STO twin walls is considerably more intricate than previously assumed. It is now understood that the simple flexoelectric picture is incomplete, and at least three distinct coupling mechanisms (Fig. \ref{Fig2}b-e) have been identified as contributing:

\begin{enumerate}
\item \textbf{Flexoelectric (rotoflexo) coupling}, which links octahedral rotations ($\phi$) to strain gradients, consistent with earlier continuum and phenomenological models.
\item \textbf{Rotopolar coupling}, an “improper” polarization mechanism involving the amplitude of the octahedral tilts ($u_{Ti}$) and their spatial gradients.
\item \textbf{A trilinear coupling} between AFD tilts and antiferroelectric displacements of Ti ions, providing an additional and previously unrecognized pathway for local polar distortion.
\end{enumerate}

On the experimental front, although considerable direct and indirect evidence supports the flexoelectric mechanism, the contributions of rotopolar and trilinear couplings remain to be conclusively verified. Another fundamental open question concerns the true ferroelectric nature of these polar twin walls, specifically, whether they can sustain switchable bipolar states? While first principle~\cite{He:2024p176801} and molecular dynamics calculations~\cite{Lu:2025p206103,Lu:2025p184105} indicate towards such a possibility by tuning flexoelectric and rotopolar coupling through stress gradients and high electric field, respectively, definitive experimental validation is still pending.

Interestingly, the emergent polarity at ferroelastic twin walls also couples strongly to the bulk order parameters of STO. Coupling to the ferroelastic order is particularly well established~\cite{Zubko:2007p1667601} and can be naturally attributed to the role of flexoelectric coupling, which itself is a primary contributor to the emergence of polarity at the twin wall. A prominent manifestation of this coupling is the electric-field-driven motion of domain walls wherein an applied electric field induces strain via electrostriction, generating strain gradients that couple back to the polar twin wall and drive its displacement (Fig. \ref{Fig4}). An important question, whether such polar twin walls can remain tunable by strain, stress, or electric fields even in the presence of electron doping will be revisited later and examined in detail. By contrast, coupling between the polar twin walls and the suppressed ferroelectric order, as well as associated quantum fluctuations near the ferroelectric quantum critical point, has only recently begun to be appreciated~\cite{Pesquera:2018p235701,Kustov:2020p016801,Fauque:2022pL140301,Casals:2019p032025}. This coupling has been shown to give rise to novel quantum phases, including the quantum domain glass and quantum domain solid phase~\cite{Pesquera:2018p235701,Kustov:2020p016801}. Interestingly, these intriguing phases remain influential even after electron doping. To preserve the logical flow of this article, we defer a detailed discussion of quantum fluctuation effects to Section 5, where we focus on the dynamical aspects of domain walls.

\begin{figure*} [t!]
 \centering
	\vspace{-1pt}
	\hspace{0pt}
	\includegraphics[width=0.65\textwidth] {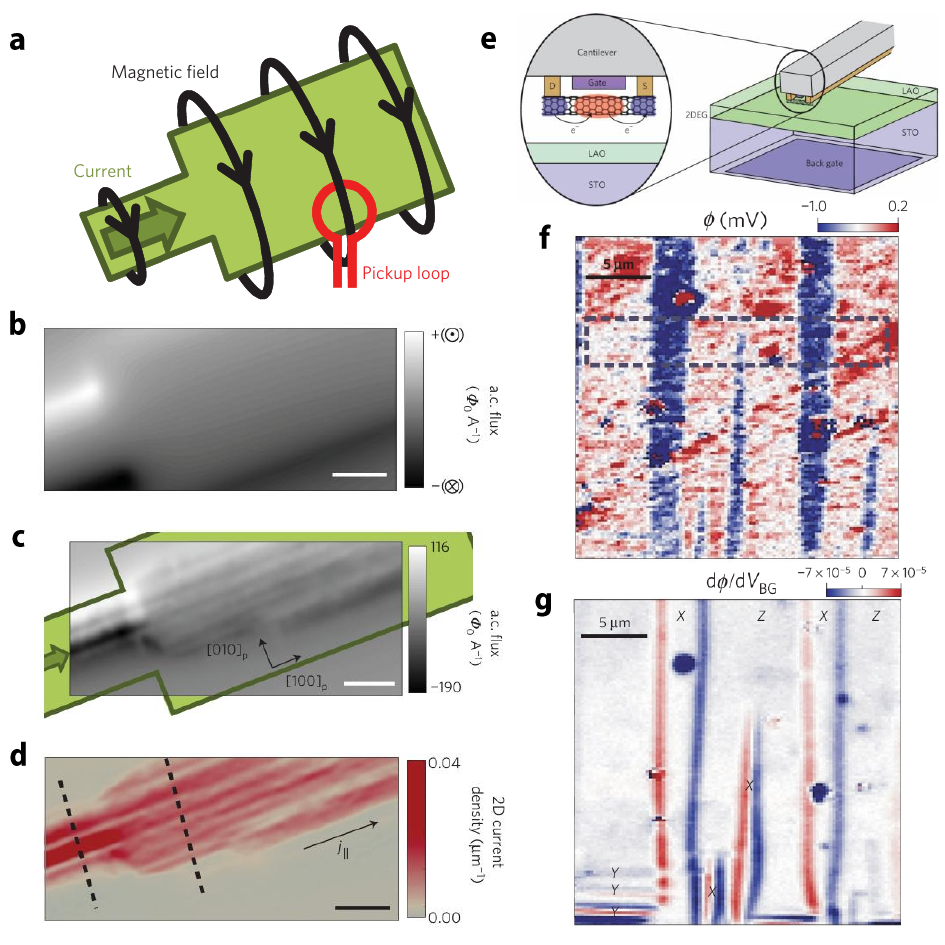}
	\caption{\textbf{a}-\textbf{d} Scanning SQUID microscopy, used to visualize local current distribution in LAO/STO heterostructures. \textbf{b}. Simulated magnetic flux map for a uniformly conducting sample. \textbf{c}. Experimental magnetic flux image reveals alternating positive and negative flux signals. \textbf{d}. Reconstructed current density map confirms that current preferentially channels along specific domain orientations emphasizing the strong coupling between electronic transport and the underlying domain structure. Panels \textbf{a}-\textbf{d} are taken from ref.~\cite{Kalisky:2013p1091} \textbf{e}-\textbf{g}. Electrostatic imaging of domain walls using a scanning single-electron transistor (SET). \textbf{e}. Illustration of the SET measurement setup. A carbon nanotube-based SET probe is scanned over the back-gated LAO/STO surface. \textbf{f,g}. Surface potential and electromechanical response maps revealing the electrostatic modulation induced by tetragonal domains in LAO/STO. Panels \textbf{e}-\textbf{g} are taken from ref.~\cite{Honig:2013p1112}.\label{Fig5} }
\end{figure*}

\begin{figure*} [t!]
	\vspace{-1pt}
	\hspace{0pt}
	\includegraphics[width=\linewidth] {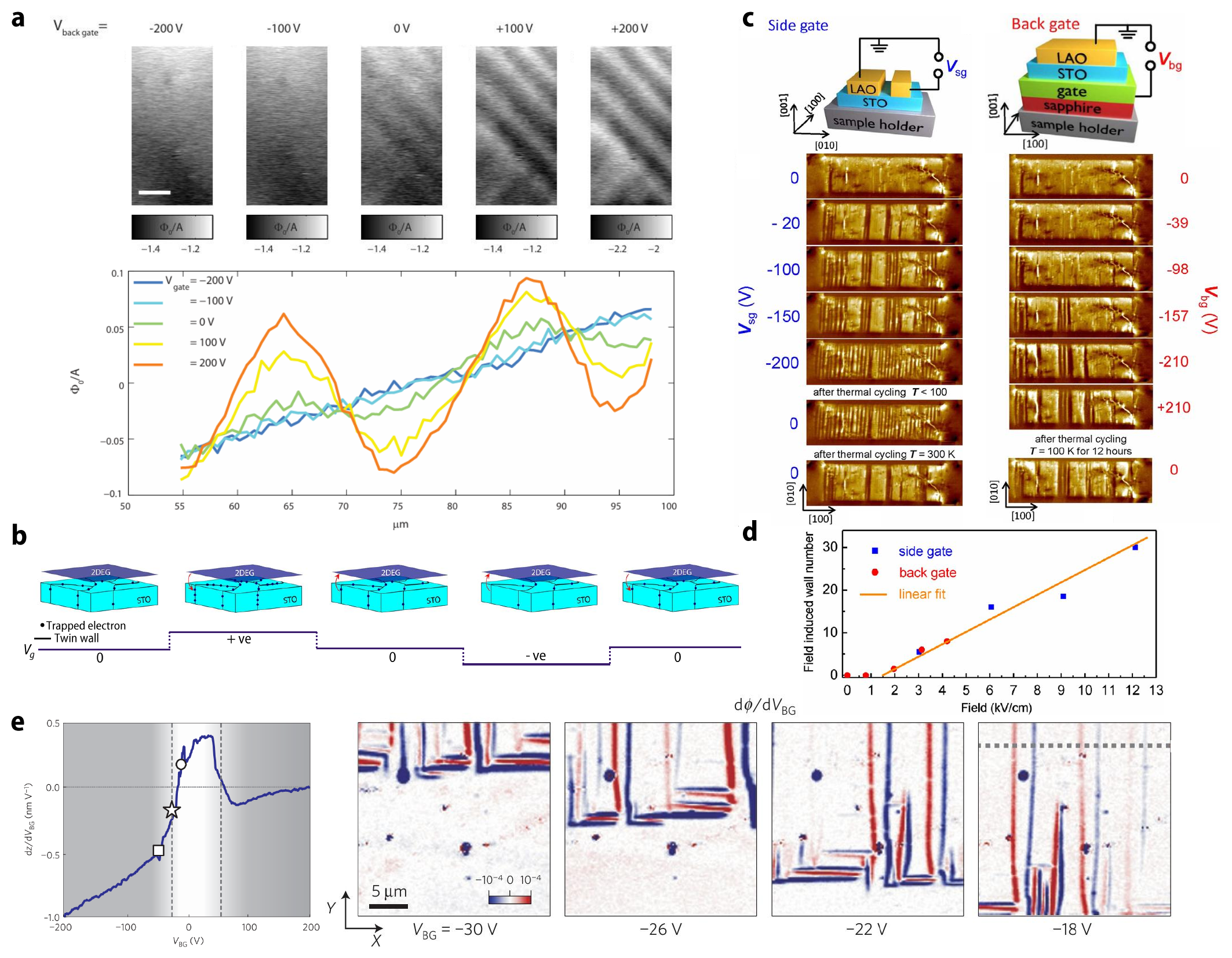}
	\caption{\textbf{a}. Gate-voltage-dependent modulation of current paths in LAO/STO heterostructure. SQUID images show that with increasing gate voltage, the stripe-like current channel becomes more pronounced, indicating enhanced conductivity. Panel \textbf{a} is taken from ref.~\cite{Kalisky:2013p1091}. \textbf{b}. Schematic illustration of electric-field-driven trapping and detrapping of charges at ferroelastic twin walls in STO, affecting interfacial transport in the 2DEG. Panel \textbf{b} is taken from ref.~\cite{ojha:2021p054008}. \textbf{c}, \textbf{d}. Gate-controlled LTSEM measurements reveal the formation and stabilization of ferroelectric twin walls in STO under applied electric fields. Both side and back-gate configurations produce similar domain wall patterns, with the onset of ferroelectric order occurring above a threshold field of around 1.4 kV/cm. Panels \textbf{c}, \textbf{d} are taken from ref.~\cite{Harsan:2016p257601}. \textbf{e}. A sequence of lateral electromechanical response maps shows gate field-induced motion of domain wall in the STO-based heterostructure. Panel \textbf{e} is taken from ref.~\cite{Honig:2013p1112}.\label{Fig6} }
\end{figure*}

\section{Static influence of Ferroelastic Domain Walls on doped electrons}

Having established the relevant bulk order parameters and the emergent symmetry breaking at ferroelastic twin walls, we now turn to the central theme of this review—how these twin walls mediate and tune the local and macroscopic responses of the system upon electron doping. To maintain conceptual clarity, this section addresses only the static influence of the twin walls—by which we mean that much of the underlying physics can be understood by considering even a single, static domain wall, without invoking interactions among neighboring walls. The dynamics of twin walls and their collective behavior will be discussed in the following section. The discussion is organized as follows. Section 4.1 focuses on the coupling between itinerant charge carriers and the emergent polar order confined at the twin walls, exploring how this coupling can be harnessed to engineer tunable functionalities.  The subsequent Sections 4.2 and 4.3 address the broader implications of ferroelastic twin walls on various charge transport phenomena. The last subsection, 4.4, highlights a couple of additional emergent electronic and magnetic orders, such as superconductivity and magnetism induced by doping and their coupling to twin wall physics.
\subsection{Electron trapping/enhanced conduction along twin wall and its tunability}

Among the most intriguing impacts of ferroelastic domain walls on electronic transport in electron-doped STO is the interactions of doped electrons with the local polar order at twin walls, which has been the focus of sustained investigation over the past decade. From a theoretical standpoint, the polar fields confined to twin walls are expected to attract charge carriers; however, in a conventional metal, such effects should be strongly suppressed by screening effects~\cite{Ashcroft:1976p,Rischau:2017p643}. Remarkably, this expectation fails in electron-doped STO, where polar domain wall effects persist despite metallicity. This unexpected behavior has stimulated considerable interest and renewed theoretical and experimental efforts, particularly in light of the growing field of polar metals and unconventional superconductors~\cite{Rischau:2017p643,Bhowal:2023p53,Wang:2019p61,Volkov:2022p4599,Yu:2022p63,Michaeli:2012p117003}. The first real-space evidence of charge trapping at domain walls in metallic STO came from two independent groups—K. Moler’s group at Stanford University~\cite{Kalisky:2013p1091}  and S. Ilani’s group at the Weizmann Institute of Science~\cite{Honig:2013p1112}—using scanning Superconducting Quantum Interference Device (SQUID) microscopy and scanning Single Electron Transistor (SET) microscopy, respectively. The scanning SQUID microscope provides a sensitive probe of local current flow by mapping the magnetic flux generated by electrical currents in the sample (Fig. \ref{Fig5}a). At low temperatures, B. Kalisky $\textit{et al.}$ observed that the conductivity in the 2DEG formed at the interface between insulating LaAlO$_3$ (LAO) and STO exhibits a pronounced striped pattern (Fig. \ref{Fig5}b,c,d). The dependence of this pattern on the thermal history indicated that the observed features originated from the underlying tetragonal domain structure of STO that forms below its structural phase transition at 105 K.

This striking observation was independently complemented by scanning SET measurements performed by M. Honig $\textit{et al.}$ The scanning SET technique probes local electrostatic properties through the charge induced on a suspended carbon nanotube quantum dot, isolated by a pair of p-n junction barriers (Fig. \ref{Fig5}e). The induced charge  ($\delta$Q) can be expressed as

\begin{equation}
\delta Q = C\,\delta\phi + \phi\,\delta C,
\end{equation}

where the first term (proportional to $\delta$$\phi$) reflects variations in the local electrostatic potential of the sample, and the second term (proportional to $\delta$$C$) arises from changes in the tip-sample capacitance due to the local surface displacement. Consequently, the scanning nanotube approach is exquisitely sensitive to the local electrostatic landscape. Spatial mapping of the surface potential revealed a striped pattern corresponding to the X, Y, and Z tetragonal domains, with potential differences on the order of a meV between distinct domain orientations (Fig. \ref{Fig5}f,g).

Together, these two pioneering studies provided compelling evidence that the emergent polarity at ferroelastic twin walls leads to electronic charge trapping~\cite{brinkman:2013p1085}. These findings now form the conceptual basis for understanding a large variety of electronic transport in doped STO at low temperatures. Now, in the following sections, we discuss in greater depth how such charge trapping, along with its associated local and global electronic properties, can be manipulated through the application of an external electric field or mechanical stress thereby also shedding light on the validity of coupling between polar order and ferroelastic order parameters upon electron doping.

\begin{figure*} [t!]
\centering
	\vspace{-1pt}
	\hspace{0pt}
	\includegraphics[width=0.66\textwidth] {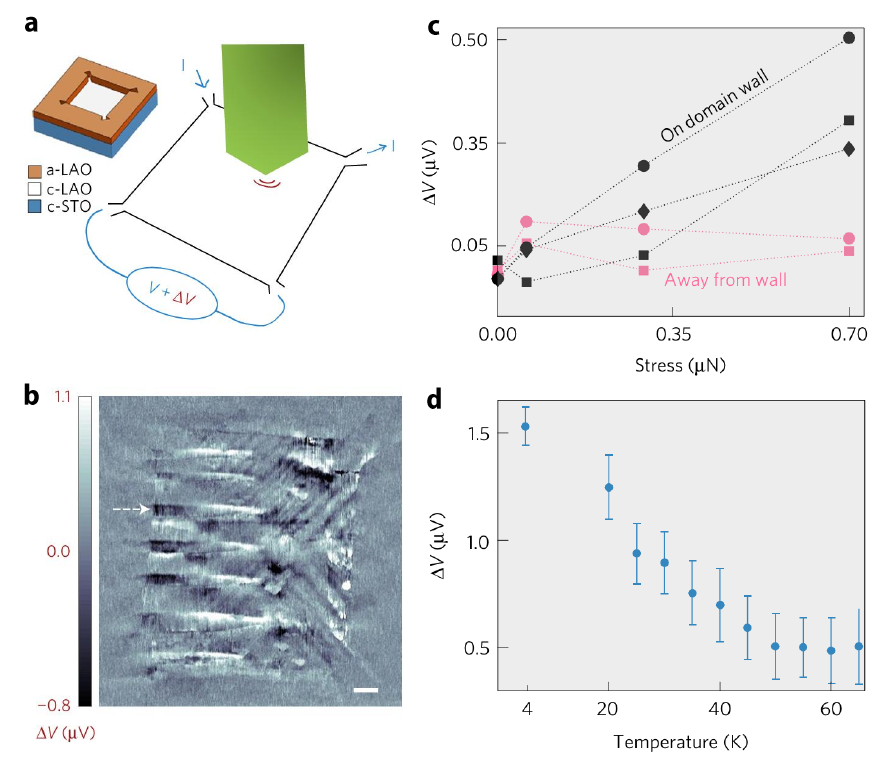}
	\caption{\textbf{a}-\textbf{d}. Visualization and control of polarity at STO domain walls using local stress. \textbf{a}. Schematic of the experimental setup of scanning stress microscopy showing voltage modulation measurements performed by applying local stress with a non-conducting tip on LAO/STO heterostructures. \textbf{b}. Spatial map of voltage response ($\Delta$V) revealing enhanced polarity localized along domain walls. \textbf{c}. Stress-dependent $\Delta$V measured on and away from domain walls, showing a strong electromechanical response confined to the walls. \textbf{d}. Temperature dependence of $\Delta$V modulation indicating a pronounced enhancement of domain wall polarity below 40 K, highlighting the emergence of low-temperature polar behavior. This figure is taken from ref.~\cite{Frenkel:2017p1203}.\label{Fig7} }
\end{figure*}

\begin{figure*} [t!]
	\vspace{-1pt}
	\hspace{0pt}
	\includegraphics[width=1\textwidth] {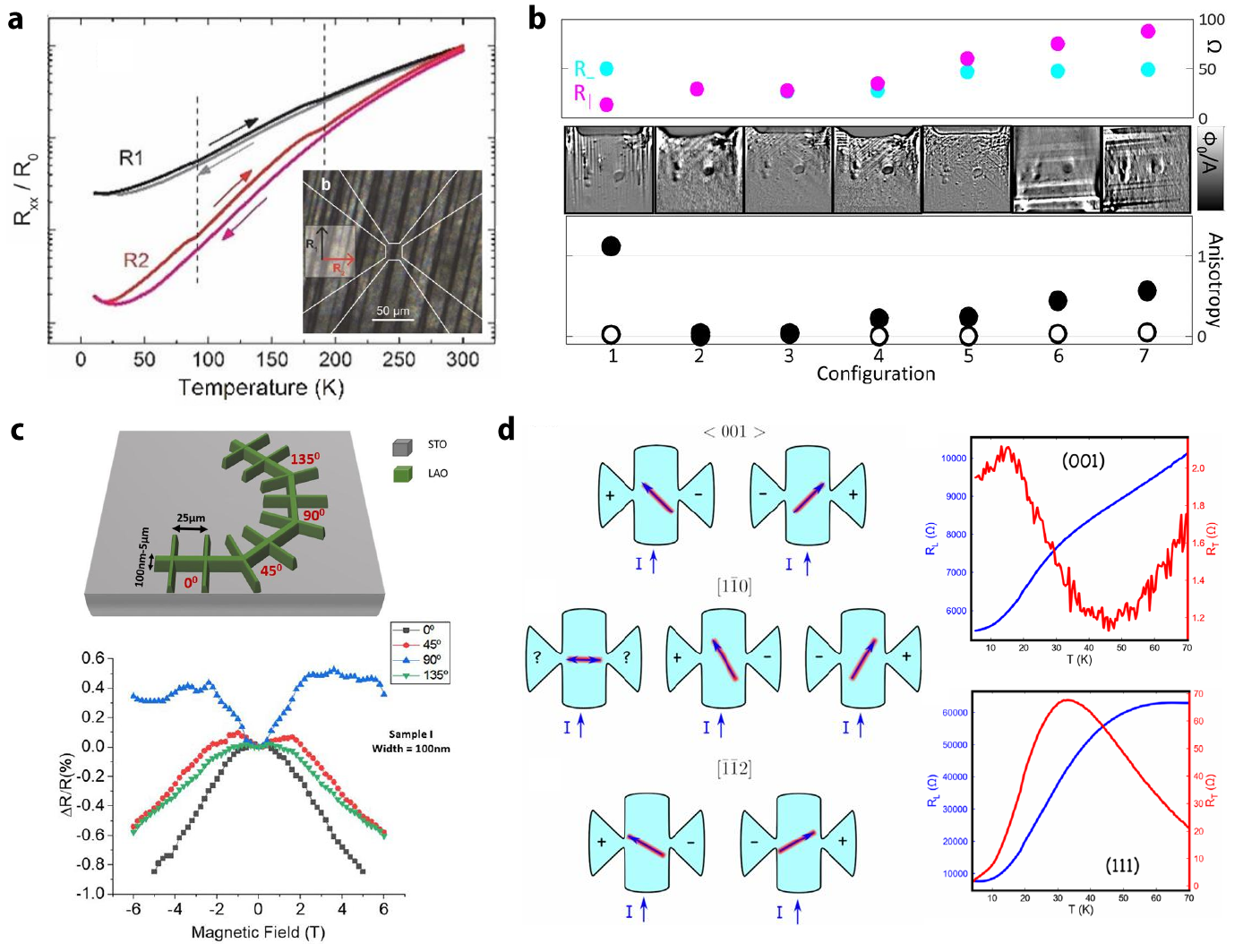}
	\caption{\textbf{a}, \textbf{b}. Temperature-dependent transport and scanning SQUID microscopy reveal anisotropic resistance in mesoscopic LAO/STO devices. Panels \textbf{a}, \textbf{b} are taken from ref.~\cite{Goble:2017p44361,Frenkel:2016p12514}. \textbf{c}. Magnetoresistance measured by varying the current direction in LAO/STO nanostructures exhibits pronounced anisotropy. Panel \textbf{c} is taken from ref.~\cite{Prasad:2021p054115}. \textbf{d}. Temperature-dependent longitudinal and transverse resistance in AlO$_x$/STO devices showing the emergence of zero-field transverse resistance below 70 K. Panel \textbf{d} is taken from ref.~\cite{Krantz:2021p036801}.\label{Fig8} }
\end{figure*}

\subsubsection{Tunability with an electric field }

While screening in a metal might be expected to suppress any electric-field control, several intriguing studies have demonstrated field-induced tunability of charge trapping at the twin wall in STO-based 2DEGs~\cite{Kalisky:2013p1091,ojha:2021p054008,Luo:2025arXiv,Lyzwa:2022p133}. B. Kalisky $\textit{et al.}$ first reported that, in the LaAlO$_3$ (LAO)/STO  system under a back-gate configuration~\cite{Kalisky:2013p1091}, the current along the domain walls increases with positive gate voltage while showing a negligible response to negative bias (Fig. \ref{Fig6}a). In sharp contrast, S. Ojha $\textit{et al.}$, studying a $\gamma$-Al$_2$O$_3$ (GAO)/STO heterostructure~\cite{ojha:2021p054008} found that the charge-trapping capacity of the twin walls could be modulated by both positive and negative gate voltages (Fig. \ref{Fig6}b). Remarkably, this tunability was found to be completely reversible down to 80~K upon removal of the electric field.

Such discrepancies likely arise from differences in the microscopic origin of the 2DEG in these systems. In LAO/STO, interfacial charge carriers are largely attributed to the polar-catastrophe mechanism~\cite{Nakagawa:2006p204,Sing:2009p176805}, whereas in GAO/STO, oxygen vacancies are the dominant source of free carriers~\cite{Schutz:2017p161409,Christensen:2017p1700026,Steegemans:2025p2119}. As oxygen vacancies are also known to impact domain wall polarity~\cite{zhang:2024p7751}, this suggests a vacancy-mediated mechanism may be crucial for the observed field response~\cite{Gunkel:2020p12,Yin:2020p017702,Hanzig:2013p024104,de:2012p174109,seri:2013p125110}. Nonetheless, this remains an open question that warrants further controlled experiments to disentangle the various contributions.

Beyond tuning the properties of existing walls, electric fields can also nucleate new conducting twin walls and induce their motion in real space~\cite{Honig:2013p1112, Harsan:2016p257601,Prasad:2023p013166,Timan:2014p082907}. In an interesting study, D. Qiu $\textit{et al.}$ found that the sketched nanowires on LAO/STO heterostructure through conductive AFM tip can seed formation and bear properties of ferroelastic domain walls~\cite{Qiu:2023p155408}. On the domain wall motion, though it is a dynamic process (discussed in Section 5), we discuss it here as a static reconfiguration in response to a field. Using low-temperature scanning electron microscopy (LTSEM), H. Ma $\textit{et al.}$ demonstrated~\cite{Harsan:2016p257601} that conducting domain walls can be nucleated above a threshold field through both back and side gating geometries (Fig. \ref{Fig6}c,d). Complementary work by M. Honig $\textit{et al.}$ on LAO/STO~\cite{Honig:2013p1112} revealed that an applied voltage can induce lateral motion of the domain walls by approximately $\sim$1 $\mu$m per volt (Fig. \ref{Fig6}e). This lateral shift corresponds to a vertical displacement of roughly 1$\sim$nm~V$^{-1}$, indicative of a giant local piezoelectric response. These findings collectively underscore that polar twin walls in STO based systems are not merely static defects but dynamically reconfigurable, electrically active entities.

\subsubsection{Tunability with stress}
The strong coupling between the polar order at twin walls and the ferroelastic order in the surrounding bulk naturally motivated investigations into how this interplay persists under electron doping. Y. Frenkel $\textit{et al.}$ provided direct experimental evidence of such coupling using scanning stress microscopy (Fig. \ref{Fig7}). They demonstrated that, in LAO/STO 2DEG system, the voltage drop across a twin wall—arising from trapped charges localized at the wall can be modulated by the application of local mechanical stress~\cite{Frenkel:2017p1203} through an atomic force microscope (AFM) tip. Remarkably, the voltage drop varied almost linearly with the applied local stress, revealing a direct electromechanical response of the wall.

 In a complementary study on another STO based 2DEG system (GAO/STO), S. Ojha $\textit{et al.}$ found that the charge-trapping capability of the twin walls also scales linearly with an applied electric field~\cite{ojha:2021p054008}. This one to one correspondence in the tunability of charge trapping with the stress and electric field strongly suggests that the ferroelastic and polar order parameters at the twin walls are intimately coupled even in the doped regime. Application of local stress can be thought of as a mechanical analogue to the electrical gating experiments. Applying local stress effectively alters the wall polarity, thereby modulating carrier accumulation or depletion and influencing the local current distribution.

 From a device perspective, electric field and stress tunability offers a tantalizing opportunity: in nanoscale structures, individual twin walls could function as mobile, spatially well-separated local gates, whose position and polarity can be controlled by external electric fields or mechanical stress. This concept opens an exciting pathway toward reconfigurable domain-wall nanoelectronics, where devices are not fixed to predefined geometries but can instead be dynamically created, erased, and tuned in situ, pointing towards possibility of polar domain wall based racetrack devices similar to the magnetic systems~\cite{Parkin:2008p190}.

\subsection{Anisotropic electron transport and zero magnetic field transverse resistance}
One of the most direct consequences of charge trapping and enhanced conductivity at ferroelastic twin boundaries in STO is their pronounced influence on electronic transport across macroscopic length scales. N. J. Goble $\textit{et al.}$ demonstrated that twin walls in the conducting LAO/STO heterostructure lead to a significant anisotropy in longitudinal resistivity~\cite{Goble:2017p44361}, depending sensitively on the orientation of the applied current direction relative to the twin boundaries (Fig. \ref{Fig8}a). Building upon this work, Y. Frenkel $\textit{et al.}$ performed a systematic investigation using scanning SQUID microscopy to directly image the current distribution in the vicinity of twin walls~\cite{Frenkel:2016p12514}. Their measurements provided a quantitative assessment of transport anisotropy for various domain-wall configurations, establishing that the enhanced conductivity along the twin boundaries acts as preferential channels for charge flow (Fig. \ref{Fig8}b).

Further insights into the role of twin walls in transport were obtained by M. S. Prasad $\textit{et al.}$~\cite{Prasad:2021p054115,Prasad:2023p013166}, who found that these conducting boundaries strongly influence the magnetoresistance (MR) in STO-based conducting nanostructures (Fig. \ref{Fig8}c). In devices with lateral dimensions below 500 nm, the MR exhibited random variations in both magnitude and sign. Crucially, upon warming the devices above $T_{AFD}$ (and thus removing the domain pattern) and re-cooling, the MR response changed significantly, including sign reversals. These results suggest that domain walls of different orientations contribute distinct magnetoresistive responses, and that the overall MR reflects the random twin-wall pattern formed during the structural phase transition upon cooldown.

An even more striking manifestation of twin-wall conduction was reported by P. W. Krantz $\textit{et al.}$ in AlO$_x$/STO heterostructures hosting a 2DEG (Fig. \ref{Fig8}d). They observed a finite transverse resistance in Hall bar devices even in the absence of an external magnetic field—a clear signature of broken symmetry in charge transport~\cite{Krantz:2021p036801}. This transverse signal appears below approximately 70~K and becomes markedly enhanced below 40~K, coinciding with the onset of polar order at the twin boundaries. Notably, the effect is far more pronounced in (111)-oriented heterostructures compared to their (001)-oriented counterparts. Capacitance measurements between the conducting interface and a back electrode revealed that this orientation-dependent behavior arises from a stronger variation of the dielectric constant with electric field along the [111] direction.

\begin{figure*} [t!]
	\vspace{-1pt}
	\hspace{0pt}
	\includegraphics[width=\linewidth] {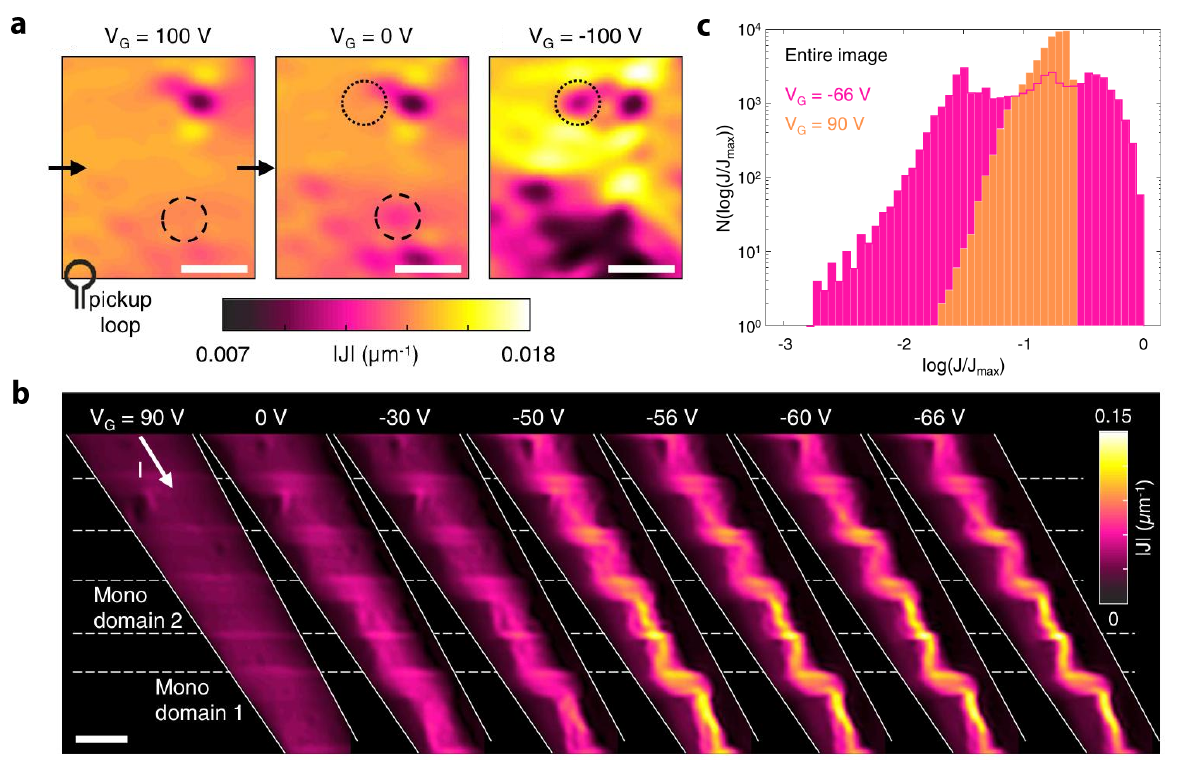}
	\caption{\textbf{a}. Scanning SQUID images showing the reconstructed current density in a monodomain region of LAO/STO heterostructure at different back-gate voltages (V$_g$). As V$_g$ is reduced, the system evolves from a uniform metallic state to a highly inhomogeneous current distribution, indicating percolative conduction near the metal–insulator transition. \textbf{b}, \textbf{c}. Current density maps and corresponding histograms in a multi-domain region reveal that decreasing V$_g$ enhances current confinement along ferroelastic domain boundaries and broadens the current distribution, indicating increased electronic inhomogeneity and breakdown of universal percolative scaling. This figure is taken from ref.~\cite{Persky:2021p3311}.\label{Fig9} }
\end{figure*}

\subsection{Influence on Metal Insulator Transition}

So far, we have discussed the static influence of domain walls on electronic transport in metallic samples. However, even in lightly doped systems—particularly those close to the metal–insulator transition (MIT), twin walls can lead to intriguing consequences~\cite{Persky:2021p3311,Bjorlig:2022p3421}. Near such transitions, macroscopic properties often obey algebraic scaling laws determined by the system’s dimensionality and underlying symmetries~\cite{Abrahams:1979p673,Lee:1985p287}. The emergence of such universal scaling signifies that microscopic details become largely irrelevant. MIT in complex oxides~\cite{Imada:1998p1039}, such as transition metal oxides, provide a fertile platform for testing the applicability of universal scaling theories. Indeed, several experiments have reported critical scaling behavior consistent with percolation models. Yet, in many of these materials, the electronic transition is entangled with other degrees of freedom—magnetic, structural, or both~\cite{middey:2016p305}. Recent work by E. Persky $\textit{et al.}$ revealed that in the presence of domain boundaries, the critical behavior of LAO/STO interfaces deviates from universality (Fig. \ref{Fig9})~\cite{Persky:2021p3311}. They observed that while a single domain follows percolation-like scaling, metallic conduction persists along domain boundaries even as the bulk two-dimensional system becomes insulating (Fig. \ref{Fig9}a). Consequently, the current-carrying backbone, defined by the specific twin-wall network rather than a random percolation path, fails to scale with the expected universal fractal dimension (Fig. \ref{Fig9}b,c). This is in sharp contrast to situation in conventional semiconductors such as Si, GaAs~\cite{Dobrosavljevic:2012p,Pollak:2013p}. Remarkably, this non-universal backbone coexists with universal scaling of the overall conductivity, and the conductivity threshold itself exhibits a size dependence. This interplay of universal and non-universal characteristics underscores the necessity of probing MIT in STO across multiple length scales to uncover their intrinsic nature.

\subsection{Emergent electronic and magnetic orders induced by doping and their coupling to twin wall physics}

Electron doping in STO introduces additional symmetry-breaking order parameters that emerge from the interplay between lattice, spin, and charge degrees of freedom. Among these, spontaneous magnetization and superconductivity represent two striking manifestations of correlated electronic/magnetic behavior in an otherwise band-insulating quantum paraelectric. The emergence of these orders, particularly near ferroelastic twin walls, adds a new dimension to the already rich landscape of competing instabilities. While the coexistence of multiple order parameters complicates the microscopic understanding of the ground state, it also presents an exciting opportunity to engineer coupled electronic/magnetic and structural phases, where twin-wall–mediated strain and polar distortions can serve as active tuning knobs for emergent magnetism and superconductivity.

\begin{figure*} [t!]
	\vspace{-1pt}
	\hspace{0pt}
	\includegraphics[width=\linewidth] {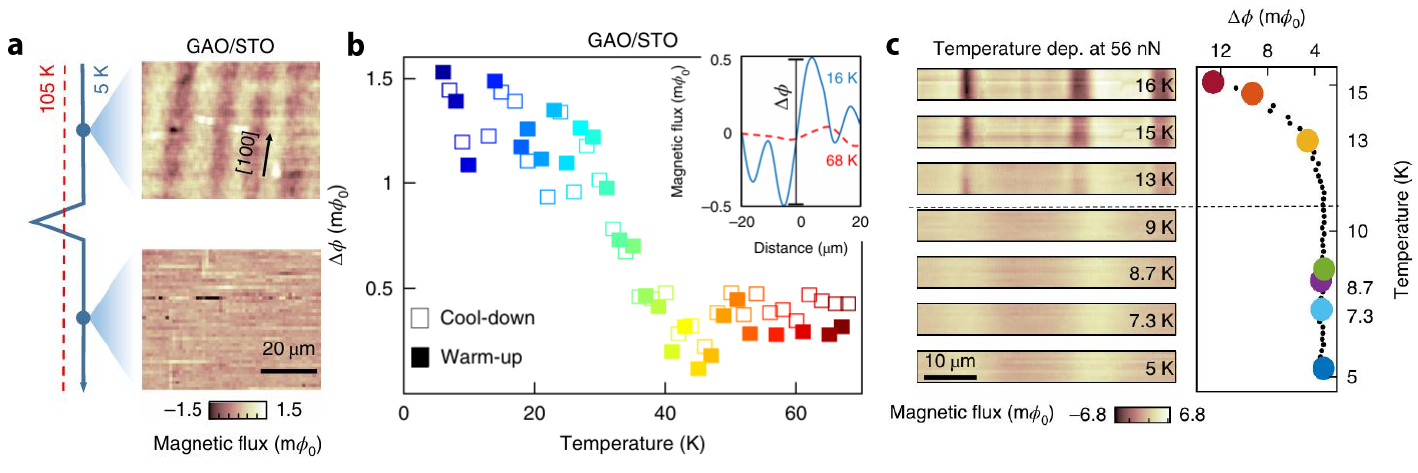}
	\caption{\textbf{a}, \textbf{b}. Temperature-dependent scanning SQUID microscopy of GAO/STO reveals that ferromagnetism emerges along ferroelastic twin boundaries below 40 K, where the local magnetic flux signal ($\Delta$$\phi$) starts to increase and becomes spatially confined to twin walls. \textbf{c}. Application of local stress modulates the magnetic flux amplitude, demonstrating strong coupling between the magnetic and ferroelastic order parameters. This figure is taken from ref.~\cite{Christensen:2019p269}.\label{Fig10} }
\end{figure*}

\begin{figure*} [t!]
	\vspace{-1pt}
	\hspace{0pt}
	\includegraphics[width=\linewidth] {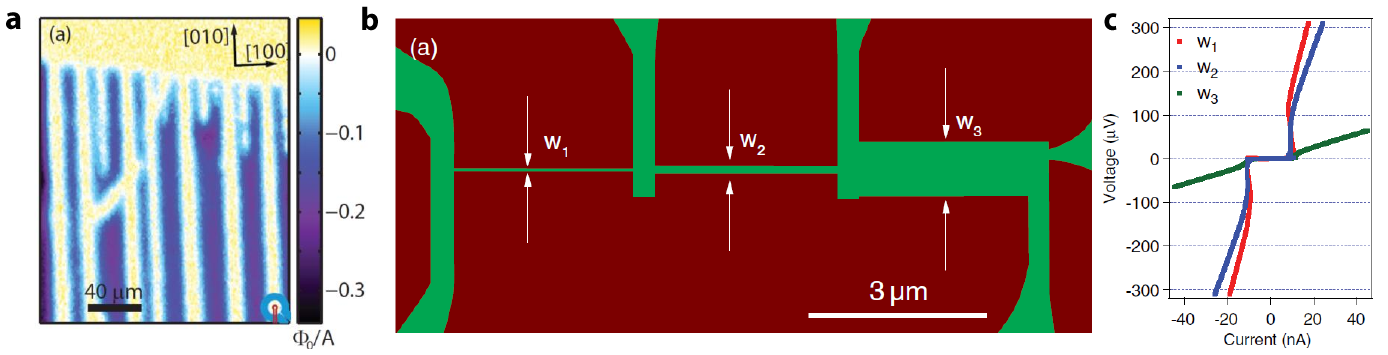}
	\caption{\textbf{a}. Scanning SQUID susceptibility maps of $\delta$-doped STO showing elongated superconducting regions aligned along the [100] and [010] tetragonal domain orientations, revealing that structural domain walls modulate the local superconducting transition temperature.  Panel \textbf{a} is taken from ref.~\cite{Noad:2016p174516}. \textbf{b}, \textbf{c}. Schematic of nanowires of varying widths (w$_1$, w$_2$, w$_3$) written at the LAO/STO interface using conductive AFM lithography, and corresponding current–voltage characteristics measured at 50 mK. The nearly identical critical currents for different widths indicate that superconductivity is confined to narrow conducting channels rather than the entire wire cross-section. This width-independent superconducting transport suggests that the supercurrent flows along ferroelastic domain walls in STO, which act as preferential one-dimensional conducting pathways at the interface. Panels \textbf{b}, \textbf{c} are taken from ref.~\cite{Pai:2018p147001}.\label{Fig11} }
\end{figure*}

\subsubsection{Magnetism and its tuning with stress}

STO, a prototypical $\textit{d}$$^{0}$ perovskite, is intrinsically nonmagnetic. However, upon electron doping, a variety of experiments have revealed the emergence of ferromagnetism on both local and macroscopic scales~\cite{Coey:2019p652,Coey:2016p485001}. The phenomenon becomes particularly striking in the context of 2DEGs formed at STO based interfaces, where ferromagnetism has been observed to coexist with superconductivity~\cite{Bert:2011p767,Li:2011p762,Dikin:2011p056802,Moetakef:2012p021014}. While considerable debate still remains whether they are phase separated in the bulk or not~\cite{Bert:2011p767,Li:2011p762,Dikin:2011p056802,mohanta:2013p025705,banerjee:2013p626}, such a coexistence would point towards an unconventional electronic ground state, where competing and coupled order parameters give rise to rich correlated behavior and potential routes toward engineered nontrivial superconducting phases.

Using scanning SQUID microscopy, D. V. Christensen $\textit{et al.}$ demonstrated~\cite{Christensen:2019p269} that apart from the bulk, ferromagnetism emerges quite robustly along ferroelastic twin boundaries (Fig. \ref{Fig10}a) in the electron-doped regime. These twin walls exhibit a pronounced enhancement of the magnetic signal below $\sim$ 40 K, coinciding with the emergence of local polar order—thereby rendering them effectively multiferroic (Fig. \ref{Fig10}b). This discovery introduces the fascinating possibility of quasi-one-dimensional multiferroicity. Interestingly, polar twin walls have also been found to significantly influence the magnetism of coupled magnetic layers, thereby allowing possibilities of engineering hybrid magnetoelectric multiferroics~\cite{Rosenberg:2017p074406,fontcuberta:2015p13784,Salje:2020p164104}. Furthermore, the intimate coupling between ferromagnetism and the ferroelastic order parameter has been directly visualized, with local magnetic moments at the walls being tunable by applied stress through an AFM tip (Fig. \ref{Fig10}c). In a further intriguing consequence, itinerant charge carriers can interact with these localized magnetic moments, giving rise to an anomalous Hall effect—a hallmark of spin-charge coupling. This interplay between free carriers and emergent magnetism underscores the potential of  STO based systems as a versatile platform for exploring spin-orbit-entangled transport and for developing oxide-based spintronic functionalities~\cite{Fert:2024p015005}.

\subsubsection{Superconductivity}

Electron-doped STO represents one of the most dilute superconductors known~\cite{Collignon:2019p031218}, with carrier densities several orders of magnitude lower than in conventional metals. Despite over half a century of study, the microscopic pairing mechanism in this system remains unresolved. Intriguingly, STO exhibits several phenomenological parallels with high-temperature superconductors: a dome-shaped dependence of the superconducting transition temperature ($T_c$) on carrier concentration~\cite{Schooley:1965p305,Collignon:2019p031218,Lin:2013p021002,Lin:2014p207002,Koonce:1967p380,GASTIASORO:2020p168107}, the presence of a low-density pseudogap-like phase~\cite{Marshal:2016p046402}, a small Fermi energy comparable to or even smaller than the Debye frequency~\cite{Eagles:1969p456,Lin:2013p021002}, and proximity to competing structural and ferroelectric instabilities~\cite{Rischau:2017p643,Edge:2015p247002}. These features have motivated a wide range of proposed pairing mechanisms primarily underscoring the complexity of electron–lattice coupling in this quantum paraelectric host.

\begin{figure*} [t!]
\centering
	\vspace{-1pt}
	\hspace{0pt}
	\includegraphics[width=.8\textwidth] {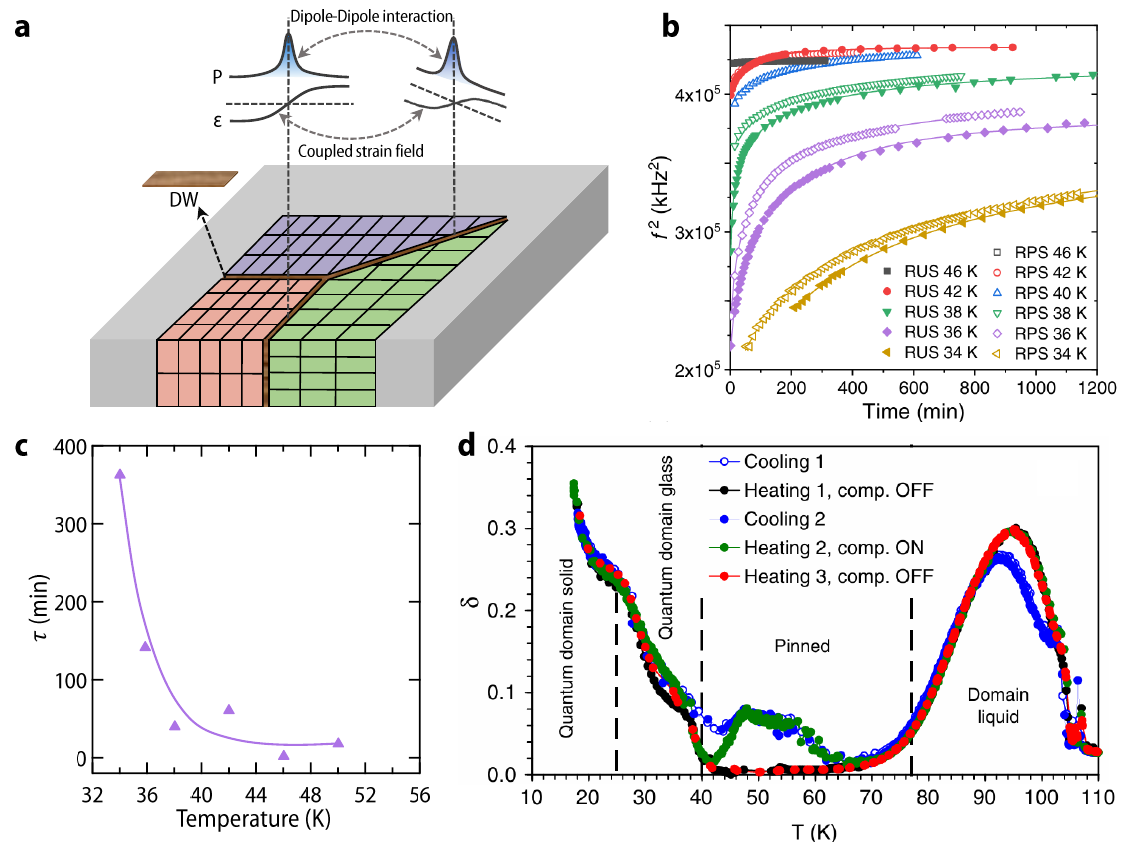}
	\caption{\textbf{a}. Schematic to depict the coexistence of strain field ($\epsilon$) and polarization ($P$) within the domain wall. A spontaneous strain field arises due to inherent ferroelasticity, which also varies in sign across the domain wall~\cite{Nataf:2020p634}. Further, the maximum value of polarization appears at the domain wall. Within a domain wall, both polarization and strain field are coupled. Additionally, one domain wall can also interact with the neighboring one via either dipole-dipole interaction or through coupled strain fields~\cite{Pesquera:2018p235701,Kustov:2020p016801}. \textbf{b},\textbf{c}. Temperature- and time-dependent resonant ultrasound spectroscopy (RUS) and resonant piezoelectric spectroscopy (RPS) measurements exhibit stretched-exponential relaxation, with the relaxation time ($\tau$) increasing sharply upon cooling, signifying the onset of glassy domain wall dynamics below 40 K. Panels \textbf{b},\textbf{c} are taken from ref.~\cite{Pesquera:2018p235701}. \textbf{d}. Temperature- and strain-dependent internal friction measurements reveal the emergence of distinct regimes of ferroelastic domain wall dynamics upon cooling in bulk STO;  from a high-temperature viscous domain-liquid phase to a pinned phase, followed by a quantum domain glass phase below 40 K, and finally a frozen quantum domain solid phase below 25 K. Panel \textbf{d} is taken from ref.~\cite{Kustov:2020p016801}\label{Fig12} }
\end{figure*}

The role of structural domains and twin boundaries in modulating superconductivity has recently come to the forefront. H. Noad $\textit{et al.}$ investigated $\delta$-doped STO using a scanning SQUID susceptometer and revealed spatial variations in $T_c$ that correlate directly with the underlying tetragonal twin structure~\cite{Noad:2016p174516} (Fig. \ref{Fig11}a). They further proposed that these modulations arise from local changes in the dielectric environment associated with the lattice orientation relative to the superconducting plane and with dielectric perturbations near twin boundaries. The influence of twin walls on superconductivity was further supported by work from Pai $\textit{et al.}$, who used conductive AFM tip to write nanoscale conducting channels at the LAO/STO interface~\cite{Pai:2018p147001}.  They found that the superconducting properties were largely independent of channel width, suggesting that electron pairing may be localized near ferroelastic twin boundaries (Fig. \ref{Fig11}b,c).

Irrespective of the exact pairing mechanism, superconductivity in the quasi-one-dimensional limit is of fundamental interest~\cite{Altomare:2013one,Oreg:2010p177002,Lutchyn:2010p077001,Vivek:2017p165117}, particularly in the context of proposals for realizing Majorana fermions in low-dimensional superconductors~\cite{Pai:2018p147001,Fidkowski:2013p014436,mohanta:2014p60001}. While direct evidence now links twin-wall structure to spatial modulation of superconductivity in STO based systems, understanding how these emergent superconducting channels respond to local mechanical stress and electric fields remains an open frontier—one that promises rich opportunities for exploring tunable quantum phenomena at the intersection of structure, polarity, and superconductivity.

Collectively, the studies reviewed above in the entire section IV establish that ferroelastic domain walls in doped STO are far from being simple structural defects. As we have seen, they act as active, quasi-1D striped patterns that locally trap charges (Sec. 4.1), break macroscopic transport symmetries (Sec. 4.2), disrupt universal scaling at the MIT (Sec. 4.3), and even provide templates for emergent correlated orders like unconventional magnetism and superconductivity (Sec. 4.4). However, all these phenomena treat the domain walls as a static individual entity. A new frontier of physics emerges when one considers that these domain walls are not fixed in time but are themselves dynamic, interacting entities, a topic we will explore next.

\vspace{0.5 cm}

\section{Complex interactions among twin walls-Introduction to dynamics}

\begin{figure*} [t!]
	\vspace{-1pt}
	\hspace{0pt}
	\includegraphics[width=\linewidth] {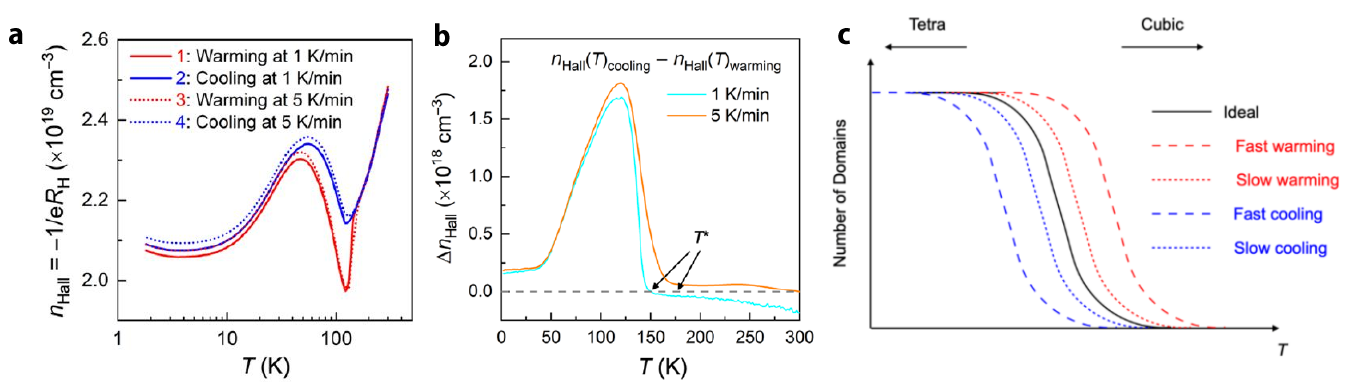}
	\caption{\textbf{a},\textbf{b}. Temperature-dependent Hall carrier density ($n_{\text{Hall}}$) of Nd-doped STO thin films measured during cooling and warming at different ramp rates reveals a pronounced hysteresis near the structural transition ($T_{AFD}$). $T^*$ indicates the onset of the difference in $n_{\text{Hall}}$ between warming and cooling cycles. \textbf{c}. Schematic illustration of the evolution of domain population in STO during temperature cycling at various ramp rates. Here, ``ideal" refers to the near-equilibrium state. This figure is taken from ref.~\cite{yue:2022peabl5668}.\label{Fig13} }
\end{figure*}

As discussed earlier, below the AFD transition the crystal accommodates three symmetry-related variants of tetragonal domains, commonly referred to as the $a$, $b$, and $c$ domains. These differently oriented domains reorganize to minimize the total elastic strain energy of the system, owing to the intrinsic ferroelastic character of the system. Upon further cooling, the onset of polarity at the domain walls introduces additional long-range dipolar interactions between domain walls, superimposed on the pre-existing elastic interactions (Fig.~\ref{Fig12}a). In sharp contrast to conventional dielectric hosts such as Si, where an isolated electric dipole induces only a local lattice polarization, the correlation length ($r_c$) of polar regions in STO is governed by the lattice polarizability~\cite{Samara:2003pR367}. This polarizability is inversely related to the TO-$\Gamma$ mode frequency $\omega_s$~\cite{Cochran:1959p412,cochran:1960p387}. As $\omega_s$ progressively softens upon cooling, $r_c$ increases substantially at low temperatures, leading to the formation of polar nanoregions (PNRs) extending over several unit cells around the polar domain walls~\cite{Sakudo:1971p851,Shirane:1967p396,Chandra:2017p112502,Ojha:2024p3830}. Consequently, dipolar interactions in STO become comparatively strong~\cite{Samara:2003pR367}.

The competition between elastic and dipolar interactions gives rise to a rich hierarchy of mesoscale domain patterns, ranging from relatively ordered configurations to highly disordered, jammed states~\cite{Casals:2019p032025}. This phenomena closely parallels the concept of jamming as a precursor to glassy dynamics observed in a wide variety of soft and hard condensed-matter systems~\cite{ediger:2000p99,sillescu:1999p81}. As the system continues to reduce its free energy, the domains-and consequently the polar domain walls-exhibit slow, time-dependent relaxation over extended timescales. The outcome is a dynamically arrested yet strongly correlated background of interacting polar domain walls, which can be dubbed as a dipole glass phase~\cite{Vugmeister:1990p993,Viehland:1991p71,Salje:2016p163}.

D. Pesquera $\textit{et al.}$ were the first to capture such glass-like dynamics of domain walls through time-dependent elastic and piezoelectric responses~\cite{Pesquera:2018p235701} in the pristine STO. The resonance frequency ($f$) in the RPS spectra (which is a fingerprint of domain wall polarity) was found to exhibit temporal relaxation under isothermal conditions. Interestingly, relaxations set in around 40 K, the same temperature where the domain wall polarity was found to enhance, and their mobility also sets in~\cite{Scott:2012p187601,Salje:2013p247603,Pesquera:2018p235701}. Moreover, since $f^2$ scales with the elastic modulus, this slow relaxation signifies a systematic stiffening of the domain structure as the mobile polar domain walls progressively freeze into a jammed configuration. This time-dependent relaxation of $f^2$ follows a stretched-exponential behavior ($\sim$ $\exp[-(t/\tau)^\beta]$), with relaxation time diverging upon cooling (Fig. ~\ref{Fig12}b,c). Such behavior is a hallmark of glassy systems~\cite{du:2007p111,vidal:2000p695}, indicating that domain walls do not relax independently but rather as part of a strongly interacting and disordered network~\cite{Amir:2011p235,Pollak:2013p}.

 \begin{figure*} [t!]
	\vspace{-1pt}
	\hspace{0pt}
	\includegraphics[width=0.9\linewidth] {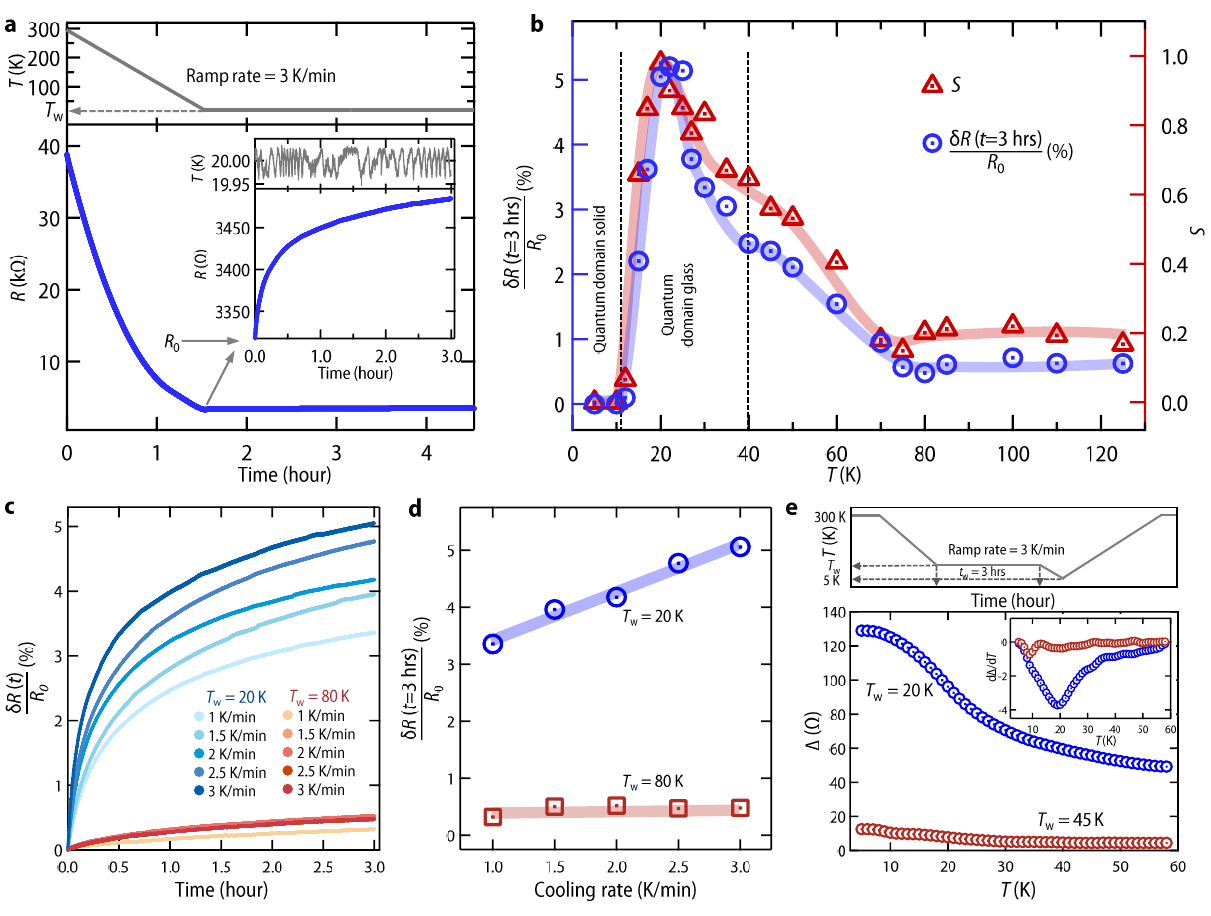}
	\caption{\textbf{a}. Protocol for measuring the temporal evolution of resistance in GAO/STO heterostructure at fixed temperature. It shows slow relaxation of resistance. \textbf{b}. The relative percentage change of resistance at the end of three hours $\frac{\delta R \textnormal{(t=3 hrs)}}{R_0}(\%)$ and relaxation rate ($S$ =$ \frac{d\left[\frac{\delta R(t)}{R_0}(\%)\right]}{d\ln(t)}$) at several fixed temperatures. This temperature dependence shows a one-to-one correspondence with the bulk STO domain wall phase diagram. The domain glass phase emerges below 40 K and persists down to 12 K before freezing into a domain solid phase. \textbf{c} - \textbf{e}. Cooling rate-dependent resistance relaxation and pronounced memory effects provide clear evidence of glassy domain wall dynamics below 40 K.  \label{Fig14} }
\end{figure*}

In another study, Kustov $\textit{et al.}$~\cite{Kustov:2020p016801} have further mapped out a complete temperature-dependent phase diagram (Fig.~\ref{Fig12}d) of the domain walls by directly measuring the internal friction ($\delta$) from the acoustic response of resonant standing waves under applied strain ($\varepsilon_0$). Their analysis revealed distinct temperature-dependent dynamical regimes. At higher temperatures, domain walls exhibit linear dynamics corresponding to a ``domain liquid” phase. Upon cooling below nearly 70 K, the motion becomes jerky, and a pinned phase appears that persists down to 40 K. Below this temperature, the non-linear jerky motion transforms into a smoother non-linear response that persists up to 25 K, marking the onset of glassy domain wall dynamics~\cite{DAnna:1997p5983}. In this regime (25 K-40 K), the internal friction follows a power-law relation with the strain, $\delta(\varepsilon_0) \propto (\varepsilon_0)^{\mu}$ ($\mu$ is the glassy exponent), which is a signature of glassy behavior. The existence of a memory effect in Young’s modulus, where the elastic response depends on the sample’s thermal history, further supports the glassy nature of domain wall dynamics (see ref. ~\cite{Kustov:2020p016801} for a detailed discussion). Interestingly, 40 K also roughly marks the onset of the quantum paraelectric regime, where quantum fluctuations assume prime importance in governing the physical properties of STO. In the light of this, the collective domain wall phase in the temperature range 25 K- 40 K has been termed as ``quantum domain glass" phase. This is a bit nontrivial, as quantum fluctuations are generally considered a bottleneck for the formation of glassy phase~\cite{Pastor:1999p4642,markland:2011p134} and therefore calls for more controlled investigations as a future research direction.

Below 25 K, the memory effect vanishes, and these non-linear domain wall dynamics start to switch to linear dynamics, accompanied by a reduction in $\mu$, which characterizes the distribution of activation energies. This reduction decreases and eventually vanishes as $T$ $\to$ 0, marking the emergence of a ``quantum domain solid phase", characterized by less glassiness, higher
stiffness, nearly linear domain wall dynamics, and frozen
domain wall relaxation. Such behavior resonates with the notion of a coherent quantum state in STO first proposed by M\"uller $\textit{et al.}$, emphasizing its proximity to a ferroelectric quantum critical point~\cite{muller:1991p277}. These results highlight how the interplay of domain wall interaction and quantum fluctuations lead to the emergence of distinct low-temperature domain wall phases~\cite{Pesquera:2018p235701,Kustov:2020p016801,Fauque:2022pL140301}, which would be extremely important for understanding electron dynamics upon electron doping in the upcoming section.

The experimental results reviewed thus far implicitly involve strain as a central perturbation, emphasizing the importance of elastic coupling in shaping collective domain-wall behavior. We next address the role of dipolar interactions, which can be more directly probed through experiments performed in the presence of an external electric field.  RUS measurements have demonstrated that even relatively weak electric fields can progressively reorganize the domain-wall network over time at fixed temperature, without modifying the underlying relaxation mechanism~\cite{Pesquera:2018p235701}. This indicates that in the linear response regime, the electric field can perturb the free-energy landscape without altering the intrinsic kinetic constraints governing domain wall motion. Subsequent optical microscopy studies revealed that, below approximately 40 K, electric fields can drive the collective motion of ferroelastic domain walls through jumps, avalanche-like events, rather than continuous smooth movement, indicating strongly correlated and jerky dynamics~\cite{Casals:2019p032025} akin to other driven glassy systems.

\begin{figure*} [t!]
	\vspace{-1pt}
	\hspace{0pt}
	\includegraphics[width=\linewidth] {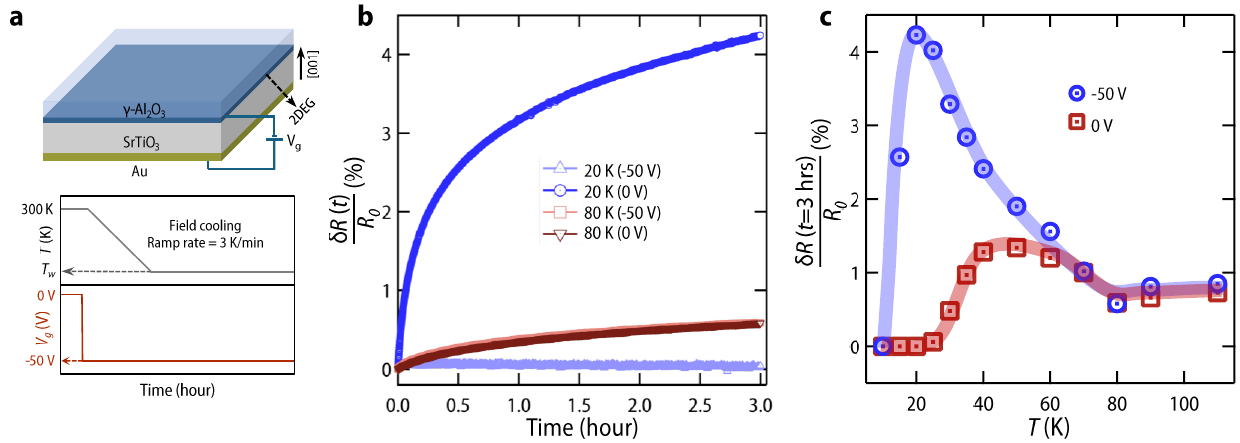}
	\caption{\textbf{a}. The top panel shows the device geometry for applying back gate voltage ($V_\textnormal{g}$). For gate field-dependent measurements, 80 nm Au layer is sputtered as a back electrode on the bottom side of the STO substrate. The bottom panel shows the temperature ramping protocol for the resistance relaxation measurement for field cooling. For zero field cooling, the sample is cooled without any back gate voltage as shown in Fig.~\ref{Fig14}a, and for field cooling, the system is cooled down from room temperature in the presence of a back gating voltage ($V_g$), -50 V to a fixed temperature ($T_w$) ) at which the temporal evolution of resistance is measured.  \textbf{b}. Temporal evolution of resistance at fixed temperatures of 20 K and 80 K after zero-field cooling and field cooling. \textbf{c.}  The relative percentage change of resistance at the end of three hours $\frac{\delta R \textnormal{(t=3 hrs)}}{R_0}(\%)$ for several fixed temperatures for zero field cooling (blue) and field cooling (red). \label{Fig15} }
\end{figure*}

\section{Glasslike electron dynamics}
The studies discussed above reveal that ferroelastic domain walls in pristine insulating STO are not static structural entities but dynamically active defects that exhibit collective motion and glasslike relaxation below 40 K. Such a complex landscape of slowly evolving, correlated domain structures naturally raises an important question of whether these dynamics would also influence the conduction electrons in metallic STO. Indications of such a coupling have been indeed observed. For instance, recent transport studies by J. Yue $\textit{et al.}$ on Nd-doped STO thin films revealed anomalous transport and thermal hysteresis in the carrier density ($n_{\text{Hall}}$) near $T_{AFD}$ (Fig. \ref{Fig13}a,b)~\cite{yue:2022peabl5668}. The hysteresis was found to be strongly ramp-rate dependent, suggesting that the behavior is not a simple electronic transition but is linked to a dynamic structural process. The proposed origin is a kinetic lag of the domain wall network, which acts as a carrier sink, behind the temperature sweep. This happens because the physical process of forming walls (during cooling) or the annihilation of them (during warming) is slow, while the temperature change (the ramp rate) is fast (Fig. \ref{Fig13}c). The slow structural changes lag behind the fast temperature ramp. This kinetic lag results in a different number of domain walls and, therefore, a different number of trapped carriers at the same temperature, depending on whether the sample is being cooled or warmed. However, since these are time-averaged measurements, they capture the cumulative impact of the domain kinetics rather than measuring the dynamics in real time.

The explicit time-dependent nature of this interplay in the low-temperature regime has been directly addressed by J. Maity \textit{et al.} through systematic investigations of electronic transport in a $\gamma$ $Al_2$$O_3$/STO heterostructure~\cite{Maity:2025p84} which hosts interfacial 2DEG as discussed earlier. In this system, the 2DEG resides within a few nanometers of the substrate surface~\cite{Sing:2009p176805,Schutz:2015p165118,Mukherjee:2016p245124,Gabel:2023p045125} and therefore 2DEG is in immediate proximity to the domain-wall network, the walls act as a dynamic, correlated disorder landscape. While static effects of domain walls, like resistive hysteresis near 90 K~\cite{Goble:2017p44361,Minhas:2017p5215,Kwak:2022p6458} and carrier freezing below $T_{AFD}$~\cite{Han:2025p191602}, are observed, the most striking results emerged in the time domain.

Upon cooling to a fixed temperature, the resistance exhibits a slow, non-saturating relaxation (Fig. \ref{Fig14}a), akin to the behavior seen in conventional electron glasses~\cite{Dobrosavljevic:2012p,Pollak:2013p,Amir:2011p235,Lee:1985p287}. Notably, the temperature dependence of this relaxation magnitude mirrors the domain-wall phase diagram of bulk STO~\cite{Kustov:2020p016801}, indicating that electron transport is intimately coupled to the underlying domain dynamics (Fig. \ref{Fig14}b).  Focusing on the regime below 40 K, strong history dependence was observed: resistance relaxation slows with reduced cooling rate at 20 K but remains unaffected at 80 K, confirming glass-like kinetics (Fig. \ref{Fig14}c,d). A memory effect measurement further confirmed this non-ergodic behavior below 40 K (Fig. \ref{Fig14}e).  These results collectively demonstrate that the domain glass dynamics are present in the metallic system below 40 K, persisting down to  $\sim$ 12 K, the onset of the quantum domain solid phase in this system, a state where this relaxation is suppressed completely (Fig. \ref{Fig14}b). Further, from the resistance relaxation at different fixed temperatures under an applied electric field, it was observed that the electric field has a negligible influence on resistance relaxation above 60 K (Fig. \ref{Fig15}). However, below 40 K (which roughly marks the onset of the ``quantum domain glass" phase) the electric field drastically suppresses resistance relaxation compared to zero-field cooling. This is consistent with the fact that applying an external electric field is known to align local dipoles and suppress randomness in dipolar glasses~\cite{Samara:2003pR367}. These observations establish a direct link between dipolar domain-wall dynamics and slow electronic relaxation, revealing a new emergent electronic state, a dipolar glassy metal in STO-based 2DEGs.

The observed resistance relaxation in the dipolar glass phase is proposed to be a direct consequence of the non-equilibrium nature of the background dipolar glass, though the precise mechanism may involve different possibilities. One scenario is that, in the glassy phase, the domain walls are considered to have an effective temperature higher than the lattice and are unable to equilibrate~\cite{Leuzzi:2007thermodynamics,Leuzzi:2009p686,Mauro:2009p75}. Over time, these walls slowly relax toward thermal equilibrium, causing their local polarization to gradually increase. As the charge-trapping ability of the walls is directly linked to their polarity, this temporal evolution leads directly to the slow, time-dependent increase in carrier trapping, leading to the observed resistance relaxation. Alternatively, the glassy dynamics may be intrinsic to the electron-dipole glass interaction itself, preventing equilibration of the electronic system even in the presence of itinerant electrons. While these scenarios provide possible frameworks for understanding non-trivial electronic relaxation in doped STO, the underlying mechanisms remain poorly understood and call for further systematic investigation in the future.

\section{Concluding Remarks and Outlook}

\begin{figure*} [t!]
	\vspace{-1pt}
	\hspace{2pt}
\includegraphics[width=0.8\linewidth] {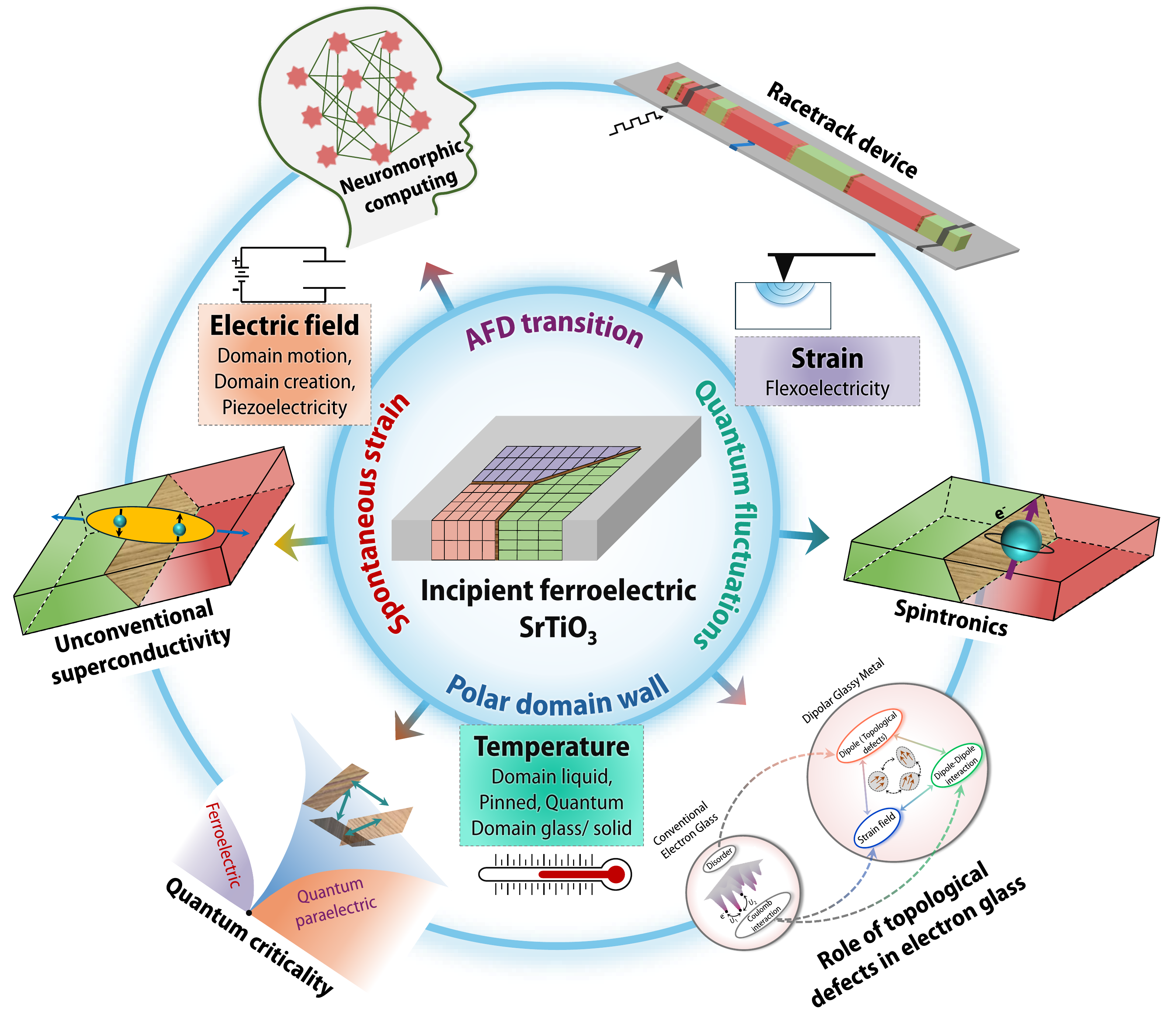}
	\caption{The central schematic shows the intrinsic landscape of STO below its antiferrodistortive (AFD) transition, where the interplay of spontaneous strain, polar domain walls, and quantum fluctuations creates a complex, multi-order-parameter system. These domain walls act as dynamically active defects that can be deterministically manipulated using three primary control knobs: electric field (driving domain motion, creation, and piezoelectricity), temperature (tuning the domain walls through domain liquid, pinned, quantum domain glass, and quantum solid phases), and strain (exploiting flexoelectric coupling). The precise control of these knobs opens diverse avenues for future research and application (outer circle), including domain wall-based memory storage (racetrack device), neuromorphic computing, oxide spintronics, the exploration of quantum criticality, and unconventional superconductivity enhanced at the domain wall. Finally, a key fundamental aspect lies in understanding the role of topological defects in electron glass physics. In a conventional electron glass, glassy dynamics arise from the competition between strong disorder and Coulomb interactions among localized carriers. In contrast, in the case of dipolar glassy metal, polar regions around electrical dipoles created near topological defects (domain walls) effectively act as a disorder, whereas dipole-dipole interaction or strain fields would assume the role of effective interaction.\label{Fig16} }
\end{figure*}

In this Review, we have provided a comprehensive overview of how doped electrons interact with the complex ferroelastic domain-wall landscape of STO below its structural phase transition, giving rise to emergent phenomena and novel functionalities across both local and global length scales. A growing body of experimental and theoretical works have established that the coupling between multiple order parameters plays a decisive role in shaping the static structure and dynamic response of these domain walls, and in turn governs the behavior of doped charge carriers. Despite these advances, significant open questions remain concerning the underlying microscopic mechanisms—particularly those related to collective dynamics and nonequilibrium phenomena in coupled domain-wall–electron systems. In the following, we highlight several of these outstanding issues and outline selected future directions within the broader interest of the condensed matter and materials science research community (Fig. \ref{Fig16}).

One of the longstanding challenges in condensed-matter physics is to identify the true ground state and the nature of low-energy excitations in systems where strong disorder and strong interactions coexist. In the strongly localized regime, such as lightly doped insulators, their subtle interplay gives rise to glassy electronic dynamics~\cite{Dobrosavljevic:2012p}. By contrast, in the opposite limit of highly delocalized electrons—namely, in conventional metals—efficient electronic screening strongly suppresses both electron–electron and electron–impurity interactions. As a consequence, these systems typically possess a nondegenerate ground state with a well-defined Fermi surface, rendering glassy dynamics largely incompatible with metallic behavior. Against this backdrop, the observation of a dipolar glassy metallic phase in doped STO represents a striking departure from the conventional electron-glass paradigm. It further suggests that, in systems such as STO, glassy electronic freezing may not act as a precursor to the MIT. This behavior stands in sharp contrast to that observed in conventional doped semiconductors such as Si, GaAs etc~\cite{Lee:1985p287,Dobrosavljevic:2012p,Amir:2011p235,Pollak:2013p}, and places STO beyond the standard Anderson and Mott localization scenarios.

While analogies can still be drawn regarding the roles of disorder and interactions in doped STO (Fig. \ref{Fig16}), it is evident that these anomalous behaviors originate from the presence of multiple, strongly coupled order parameters—an essential ingredient absent from traditional electron-glass phenomenology. Recent advances in the study of structural glasses~\cite{baggioli:2021p015501,baggioli:2023p2956} have highlighted the central role of topological defects in explaining some of their unusual properties. In this broader context, ferroelastic twin walls, as intrinsically occurring topological defects, provide a unique and previously unexplored platform for investigating the role of topological defects in electron-glass physics.

Another unresolved puzzle concerns the disappearance of glassy electronic relaxations at temperatures even lower than those associated with the established domain-glass to domain-solid transition. At first glance, one might question whether oxygen vacancies themselves play a role in this behavior. However, the problem is considerably more complex, given that the very nature of the transition from a domain-glass to a domain-solid state remains poorly understood~\cite{Kustov:2020p016801}. Resolving this inconsistency will require a more careful examination of the role of quantum fluctuations, motivating further theoretical modeling and targeted experimental investigations. An additional aspect that has been largely absent from existing discussions is the possible role of other lattice degrees of freedom. In particular, it is well known that coupling between TA phonons and polar order can give rise to polarization density waves~\cite{Orenstein:2025p961}. Whether such phonon-mediated correlations contribute to the stabilization of the domain-solid phase or to the suppression of glassy dynamics remains an open question. Furthermore, the implications of these collective lattice modes for doped charge carriers—and their coupling to the polar background—have yet to be explored and may prove crucial for understanding transport anomalies in this regime. More broadly, the questions raised above lie at the core of understanding charge transport in an emerging class of materials known as quantum-critical polar metals~\cite{Bhowal:2023p53,Ojha:2024p3830,Rischau:2017p643,Wang:2019p61,Engelmayer:2019p195121}. In this context, the observation of glassy electronic relaxations in doped STO may add to the list of key organizing principles for developing a unified theoretical framework describing conduction in quantum-critical polar metals and polar superconductors~\cite{Volkov:2022p4599,saha2025:p82}.

Another important aspect that warrants further investigation is the observation of enhanced electron pairing along ferroelastic domain walls, while the surrounding bulk remains relatively non-superconducting. Because the carrier density along the domain wall can differ substantially from that of the bulk, this raises fundamental questions about the true superconducting $T_c$ dome as a function of doping in such spatially inhomogeneous systems. Addressing this issue will require spatially resolved probes of the superconducting gap across samples with systematically varied carrier densities, in order to directly track the evolution of pairing strength and $T_c$ both along the domain walls and in the bulk. Beyond elucidating the nature of unconventional superconductivity in these materials, the existence of quasi-1D superconducting channels along with ferromagnetism opens exciting opportunities to explore emergent low-dimensional phenomena, including the realization of exotic quasiparticles such as Majorana modes and other topologically protected edge states, which hasn't been demonstrated so far in doped STO~\cite{Pai:2018p147001,Fidkowski:2013p014436,mohanta:2014p60001}.

Having outlined several fundamental open questions central to understanding the origin and consequences of coupled electron–domain wall systems, we conclude by highlighting key perspectives from a device standpoint. Firstly, since individual twin walls can function as mobile, spatially well-defined local gates whose position, polarity, and functionality can be controlled by external electric fields or mechanical stress. This capability opens a compelling route toward spatially reconfigurable domain-wall nanoelectronics, in which device elements are not constrained to fixed geometries but can instead be dynamically written, erased, and tuned in situ. Such a paradigm naturally points toward the exploration of polar domain-wall-based racetrack devices, conceptually analogous to the magnetic racetrack memory proposed by S. Parkin~\cite{Parkin:2008p190,Parkin:2015p195,Blasing:2020P1303}, but enabled here through electric-field control of polar order.

Neuromorphic computing represents another exciting frontier in modern electronics. This paradigm enables the simultaneous processing and storage of information, unattainable in present digital computers~\cite{Mondal:2024p4,Kumar:2022p575,Yuan:2024p9733,Kudithipudi:2025p801,Christensen:2022p022501}. The intrinsic properties of the STO domain-wall system highlighted in this review, specifically the strong history dependence and gradual relaxation response under perturbations such as electric fields and stress, render these materials uniquely suitable for this purpose. The slow relaxation in response to the perturbation mimics short-term synaptic plasticity, essential for temporal processing, while the ferroelectric nature of domain walls under an electric field enables long-term plasticity, required for learning~\cite{Oh:2019p15733,Chen:2025p53875}.  Consequently, these coupled electron-domain wall systems offer a promising physical platform for realizing energy-efficient neuromorphic computing architectures.

An equally intriguing direction concerns the potential utilization of confined ferromagnetism at polar domain walls. Whether such emergent magnetism can be harnessed to form quasi-1D spin channels, and whether ballistic or low-dissipation spin transport can be achieved along these channels, remains an open and technologically significant question. Addressing these challenges would not only deepen our understanding of domain wall-mediated electronic and spin phenomena, but could also establish polar domain walls as versatile building blocks for future spin-based and low-power electronic devices~\cite{Fert:2024p015005}.

\section{Acknowledgment}
SM acknowledges funding support from a SERB Core Research grant (Grant No. CRG$/$2022$/$001906) and I.R.H.P.A Grant No. IPA$/$2020/000034. JM acknowledges UGC, India for fellowship.


%

\end{document}